\documentclass[10pt,conference,letterpaper]{IEEEtran} 

\usepackage[T1]{fontenc}

\usepackage{mathptmx}
\usepackage[scaled=0.92]{helvet}
\usepackage{courier}
\usepackage{amsmath,amsfonts,amssymb,mathrsfs}
\usepackage{graphicx}
\usepackage{subcaption}
\usepackage{float}
\usepackage{algorithm,algpseudocode}
\usepackage{booktabs,array,tabularx,multirow,makecell}
\usepackage{url}
\usepackage{cite}
\usepackage[hyphenbreaks]{breakurl}
\usepackage{xcolor}
\usepackage[normalem]{ulem}
\usepackage{enumitem}
\usepackage{hyperref}
\hypersetup{colorlinks=true,linkcolor=blue,citecolor=blue,urlcolor=blue}

\newcommand{\projname}{\emph{ClawGuard}}
\newcommand{\para}[1]{\smallskip\noindent\textbf{#1}~}

\definecolor{newred}{rgb}{0.78,0.10,0.13}
\definecolor{fixblue}{rgb}{0.10,0.35,0.78}
\definecolor{cutgrey}{rgb}{0.45,0.45,0.45}
\definecolor{revgreen}{rgb}{0.05,0.45,0.10}
\DeclareRobustCommand{\new}[1]{\textcolor{black}{#1}}

\let\origsout\sout
\renewcommand{\sout}[1]{\textcolor{cutgrey}{\origsout{#1}}}
\newenvironment{newblk}{\begingroup\color{black}}{\endgroup}

\graphicspath{{fig/}}
\IEEEoverridecommandlockouts
\begin{document}

\title{\projname: Out-of-Band Detection of LLM Agent Workflow Hijacking via EM Side Channel}

\author{%
  \IEEEauthorblockN{%
    Leo Linqian Gan, 
    Jeffery Wu, 
    Longyuan Ge, 
    Lanqing Yang\textsuperscript{*}, 
    Yonghao Song, \\
    Jingkai Zhang, 
    Haojia Jin, 
    Weiyi Wang, 
    and Guangtao Xue\textsuperscript{*}\thanks{\textsuperscript{*}Corresponding authors.}
  }
  \IEEEauthorblockA{%
    Shanghai Jiao Tong University, Shanghai, China \\
    Email: \{leo-gan, jeffery2019, gly2000, yanglanqing, songyonghao, \\
    jinhaojia, weiyi\_wang, gt\_xue\}@sjtu.edu.cn, zhangjingkai05@gmail.com
  }
}
\maketitle
\pagestyle{plain}

\begin{abstract}
Autonomous Large Language Model (LLM) agents increasingly execute
state-changing workflows by invoking files, databases, shells, network services,
and external tools.  This autonomy introduces a new integrity risk:
\emph{workflow hijacking}, in which an adversary inserts, omits, reorders, or
substitutes skill invocations while preserving plausible high-level semantics.
Existing defenses rely largely on host-internal telemetry such as audit logs,
system calls, provenance graphs, or runtime monitors.  Once the agent runtime
or host OS is compromised, this telemetry shares the same trust boundary as the
attacker and can be forged, suppressed, or blinded.
We present \projname, a passive out-of-band monitor that uses electromagnetic
(EM) emanations as a software-independent physical channel for LLM-agent
workflow auditing.  The key observation is that agent skills are
seconds-scale, compositional workloads: their mixtures of computation, DRAM
traffic, storage bursts, network blocking, and idle intervals produce
macroscopic EM envelopes that can be measured by external software-defined
radios.  \projname\ converts continuous RF streams into skill-level and
fine-window physical evidence through a drift-aware coarse--fine pipeline:
dual-band SDR sensing, calibrated carrier selection, $320$-dimensional
spectral/temporal/cross-receiver features, cycle-local normalization,
temperature detrending, and record-level aggregation.
We evaluate \projname\ on a $7.82\,\mathrm{TB}$ main RF corpus containing
$12{,}232$ records across $16$ benign skills and $22$ attack skills, together
with a separate new-bands replication corpus.  On the production split,
\projname\ achieves $\mathrm{AUC}=0.9945$ and detects attacks at
$100\%$ true-positive rate with $1.16\%$ false-positive rate.  On the
survey-selected $(80,800)\,\mathrm{MHz}$ carrier pair, the same pipeline
reaches $88.3\%$ record-vote accuracy and $90.3\%$ attack recall on the
surviving attack-class subset.  These results show that passive EM sensing can
serve as a practical physical consistency check for LLM-agent workflow
hijacking under host-software compromise.
\end{abstract}


\begin{figure}[t]
  \centering
  \includegraphics[width=\linewidth]{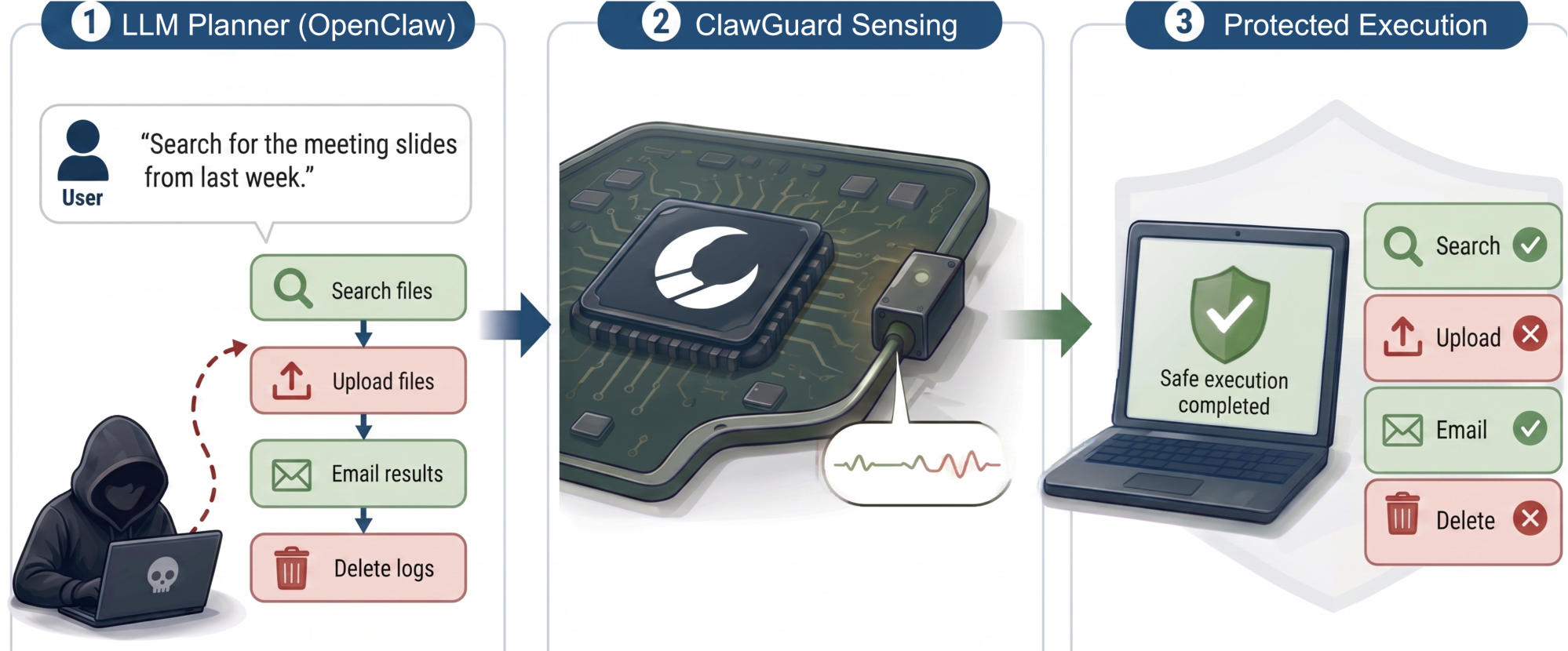}
    \caption{\new{System overview.} An adversary injects a malicious sub-skill into an OpenClaw execution on the target host, compromising host-internal telemetry. Consequently, the defender utilizes a co-located, passive SDR to observe far-field RF emanations as a secure out-of-band integrity channel.}
  \label{fig:overview}
\end{figure}

\section{Introduction}
\label{sec:intro}

As global enterprise investments in Generative AI surge toward a projected $\$140$~billion by 2030~\cite{statista_genai_2024}, the underlying computing paradigm is undergoing a profound transformation. 
Large Language Models (LLMs) are evolving from passive conversational interfaces into autonomous, task-driven agents. Frameworks like LangChain~\cite{langchain} and research vehicles like OpenClaw~\cite{OpenClaw52} integrate contextual reasoning with the ability to invoke external tools, access local file systems, and execute dynamic code. 

However, this autonomy introduces a critical vulnerability: \emph{workflow hijacking}. As enterprises deploy agents to handle increasingly sensitive financial and infrastructure tasks, the stakes of compromise have escalated. Recent security disclosures and research (e.g., PoisonedRAG~\cite{poisonedrag2025}, ToolHijacker~\cite{toolhijacker2026}, and ObliInjection~\cite{oblinjection2026}) demonstrate that adversaries can weaponize prompt injection or poisoned databases to subvert agent logic. Through these vectors, attackers can insert, reorder, or substitute tool invocations---such as injecting a covert data exfiltration branch into a routine database query---while preserving semantically plausible high-level interactions.

The conventional approach to securing such workflows relies heavily on host-internal software telemetry. To detect Advanced Persistent Threats (APTs) and workflow deviations, the security community has extensively developed system-call-based provenance graphs (e.g., HOLMES~\cite{holmes2019sp}, Unicorn~\cite{han2020unicorn}) and in-kernel eBPF observability frameworks (e.g., Kobra~\cite{kobra2023ndss}). However, these software-layer defenses suffer from a fundamental architectural flaw: a \emph{symmetric threat model}. The monitoring infrastructure shares the exact same physical and logical trust boundary as the execution environment it observes. Security audits, bootloader fuzzing campaigns~\cite{zhong2025bootloader}, and controlled preemption attacks~\cite{zhu2025controlled} consistently demonstrate that adversaries routinely achieve privilege escalation and kernel-level compromise. Once an attacker gains root access, they can effortlessly forge provenance records, blind eBPF sensors, or suppress telemetry altogether. Software cannot reliably attest to its own integrity when the underlying operating system substrate is actively controlled by an adversary. To secure high-stakes autonomous platforms, the defense mechanism must survive full host OS compromise.

To break this symmetry, workflow monitoring must be established \emph{out-of-band}, utilizing an observation channel that remains structurally isolated from the host OS. Electromagnetic (EM) side-channel signals offer such a hardware-rooted trust anchor. Arising unintentionally from the physical electrical activity of the CPU and memory bus, EM emanations reflect the ground-truth execution of the hardware. Crucially, they can be captured completely passively---without any invasive host modifications---and cannot be tampered with by software-level adversaries. 
Prior EM side-channel literature has primarily focused on bit-level cryptanalysis (e.g., RSA/ECC key extraction~\cite{genkin2014rsa,genkin2016ecdsa}) or coarse-grained application fingerprinting~\cite{sehatbakhsh2020emma}. Monitoring an LLM agent requires a distinct, intermediate granularity: the \emph{skill level}. Our key physical observation is that autonomous agent skills (e.g., database analytics, network streaming, or malicious shell executions) manifest as macroscopic, seconds-long compositional workloads. The structural choices of these skills---specifically their alternating bursts of computation, memory access, and I/O---generate highly separable, long-duration EM envelopes.

Motivated by this, we design and implement \projname, a fully out-of-band, physical-layer integrity monitor for LLM agent workflows. Deployed via software-defined radios (SDRs), \projname\ observes the host non-intrusively.
However, leveraging EM signals for semantic workflow monitoring presents a severe technical challenge: \emph{bridging the semantic gap}. Raw EM signals are noisy, continuous analog streams heavily distorted by hardware non-stationarity and thermal drift. While prior side-channel literature successfully extracts bit-level cryptographic keys~\cite{genkin2014rsa} or fingerprints coarse-grained, monolithic desktop applications~\cite{sehatbakhsh2020emma}, neither granularity suits LLM agents. Extracting high-level, discrete logical intent (e.g., distinguishing a benign file read from a malicious script execution) from continuous RF waves requires defining a completely new abstraction layer that maps macroscopic hardware physics to semantic workflows, all while remaining resilient to the temporal variance of long-running skills.

To overcome these challenges, and bridge the semantic gap between continuous analog RF streams and discrete agent intent, we introduce an event-aware coarse--fine windowing pipeline. This architecture carefully handles the substantial temporal variance and thermal drift inherent in long-running workloads. \projname\ extracts a robust feature representation to recover a physical-layer skill execution trace, which is subsequently validated against the planner's intended logic using a confusability-weighted edit distance.

Our contributions can be summarized as follows:

\begin{itemize}[leftmargin=*]
    \item \textbf{Formalizing Skill-Level Side-Channels:} We define a new mid-tier granularity for physical observation that maps macroscopic hardware emanations to semantic LLM agent workflows, establishing a hardware-rooted trust anchor capable of surviving full OS-level compromise.
    \item \textbf{Robust Out-of-Band Architecture:} We design an event-aware, drift-compensated physical-layer monitor (\projname) that bridges the analog-semantic gap without host instrumentation, effectively translating continuous RF streams into discrete workflow integrity verdicts.
    \item \textbf{Methodological Insight on RF Artifacts:} We expose a critical measurement pitfall in EM side-channel evaluations by isolating CPU-governor-driven frequency modulation from true hardware power signatures. Consequently, we provide an OS-agnostic band-selection methodology to prevent systemic evaluation artifacts in future RF research.
    \item \textbf{Large-Scale Feasibility Validation:} We validate the system on a custom $7.82\,\mathrm{TB}$ RF corpus spanning $38$ skills (including $22$ ported attack primitives). The system achieves a production-split AUC of $0.9945$ and a $100\%$ true positive rate at $1.16\%$ false positive rate with a median inference latency of $18\,\mathrm{ms}$, demonstrating the practical viability of out-of-band workflow auditing.
\end{itemize}

Our contributions can be summarized as follows:

\begin{itemize}[leftmargin=*]
    \item \textbf{Formalizing Skill-Level Side Channels:}
    We define a mid-tier physical observation granularity for LLM agents:
    the \emph{skill}.  This granularity is coarse enough to produce stable
    macroscopic EM envelopes, yet fine enough to audit workflow integrity
    beyond whole-application fingerprinting.

    \item \textbf{Out-of-Band Workflow-Integrity Architecture:}
    We design \projname, a passive, dual-SDR physical-layer monitor that
    remains outside the host software trust boundary.  \projname\ converts
    continuous RF streams into drift-compensated skill-level and attack-state
    evidence through carrier calibration, coarse--fine windowing, and
    record-level aggregation.

    \item \textbf{Large-Scale LLM-Agent RF Corpus:}
    We construct a $7.82\,\mathrm{TB}$ RF corpus for LLM-agent workflow
    monitoring, containing $12{,}232$ records across $16$ benign skills and
    $22$ attack skills, together with a separate new-bands replication corpus.
    The corpus captures synchronized dual-band IQ streams, workflow/event
    metadata, and temperature traces, enabling evaluation of skill-level EM
    separability, workflow-hijacking detection, drift, and carrier transfer.

    \item \textbf{Methodological Insight on RF Artifacts:}
    We expose a measurement pitfall in EM side-channel evaluation by separating
    CPU-governor-driven frequency modulation from workload-dependent hardware
    power signatures.  Based on a $1$--$3000\,\mathrm{MHz}$ band survey, we
    present a measured carrier-selection methodology that avoids relying on
    nominal CPU-clock or DRAM-frequency assumptions.

    \item \textbf{End-to-End Feasibility Validation:}
    We evaluate \projname\ on the main RF corpus and the new-bands replication
    corpus.  On the production split, \projname\ achieves
    $\mathrm{AUC}=0.9945$ and detects attacks at $100\%$ true-positive rate
    with $1.16\%$ false-positive rate; on the survey-selected
    $(80,800)\,\mathrm{MHz}$ carrier pair, the same pipeline reaches
    $88.3\%$ record-vote accuracy and $90.3\%$ attack recall.
\end{itemize}

\section{Preliminary}
\label{sec:background}

\subsection{Physical Basis of Workload-Dependent EM Emanations}

Electromagnetic (EM) emanations are a physical by-product of digital
computation.  In CMOS systems, instruction execution, cache activity, DRAM
transactions, I/O transfers, and clock-distribution logic all induce
time-varying current flows.  At a coarse abstraction, the dynamic power of a
digital subsystem can be written as
\begin{equation}
P_{\mathrm{dyn}}(t) \approx \alpha(t) C_{\mathrm{eff}} V(t)^2 f(t),
\label{eq:pdyn}
\end{equation}
where $\alpha(t)$ is the switching-activity factor, $C_{\mathrm{eff}}$ is the
effective switched capacitance, $V(t)$ is the supply voltage, and $f(t)$ is the
operating frequency.  Different workloads alter $\alpha(t)$ through their
instruction mix, cache behavior, memory intensity, storage bursts, network
blocking, and library calls.  On modern platforms, dynamic voltage and
frequency scaling (DVFS) may further modulate $V(t)$ and $f(t)$ in response to
sustained activity.

These current variations couple into PCB traces, power-delivery networks,
clock lines, memory buses, and attached cables, which act as unintended
radiators.  A passive receiver tuned to carrier band $f_c$ therefore observes
a noisy projection of the host's aggregate hardware activity:
\begin{equation}
X_r(t; f_c)
=
H_r(f_c)
\left(
\sum_{j \in \mathcal{C}} E_j(t; f_c)
\right)
+
N_r(t),
\label{eq:em_observation}
\end{equation}
where $X_r$ is the IQ stream captured by receiver $r$, $\mathcal{C}$ denotes
hardware components such as CPU cores, DRAM, regulators, buses, and I/O
controllers, $E_j$ is the component-specific emission, $H_r$ is the
environment- and placement-dependent channel response, and $N_r$ is receiver
and ambient noise.

Prior EM side-channel work demonstrates that such emissions can expose
cryptographic secrets at bit or operation granularity~\cite{genkin2014rsa,
genkin2016ecdsa,camurati2018screaming,wang2024bluescream}, and can also
fingerprint coarse application behavior~\cite{sehatbakhsh2020emma,
callan2014practical,yilmaz2019emi}.  \projname\ targets a different point in
this abstraction hierarchy: the \emph{skill-level} behavior of autonomous LLM
agents.

\subsection{Why Agent Skills Form a Physical Granularity}

LLM agents do not execute user requests as a single monolithic program.
Frameworks such as LangChain~\cite{langchain} and research rigs such as
OpenClaw~\cite{OpenClaw52} decompose a task into named tool invocations or
\emph{skills}.  A document-processing task, for example, may expand into
\texttt{OpenFile}$\rightarrow$\texttt{Summarize}$\rightarrow$
\texttt{ComposeEmail}$\rightarrow$\texttt{SendEmail}.  Each skill is a
bounded, seconds-scale workload that invokes host-side computation, files,
databases, shells, network services, or external APIs.

The skill abstraction is physically meaningful because it is long enough to
produce a stable macroscopic EM envelope, yet semantically fine-grained enough
to express workflow integrity.  Bit-level side channels are too fine: they
capture short cryptographic kernels rather than agent intent.  Whole-application
fingerprinting is too coarse: it cannot distinguish whether an agent performed
a benign database query, a file backup, an email send, or a malicious shell
utility within the same runtime.  Skill-level monitoring sits between these
regimes.

\begin{table}[t]
  \centering
  \small
  \caption{Granularity of EM observation. \projname\ targets the middle tier:
  coarse enough to produce stable physical envelopes, but fine enough to audit
  agent workflow integrity.}
  \label{tab:granularity}
  \begin{tabularx}{\linewidth}{lXX}
    \toprule
    \textbf{Level} & \textbf{Physical target} & \textbf{Why insufficient / useful} \\
    \midrule
    Bit / operation
      & Cryptographic rounds, key-dependent branches
      & Precise but too low-level for agent workflow semantics
        \cite{genkin2014rsa,genkin2016ecdsa} \\
    Application
      & Whole programs or user activities
      & Robust but too coarse to separate tool-level agent behavior
        \cite{sehatbakhsh2020emma,callan2014practical} \\
    Skill
      & Seconds-scale tool invocations
      & Matches the unit whose identity and order determine workflow integrity \\
    \bottomrule
  \end{tabularx}
\end{table}

For a skill $s$, we view its physical manifestation as a distribution over
windowed EM features:
\begin{equation}
s \;\mapsto\; \mathcal{P}_s
=
\Pr\!\left[
\phi\!\left(X_{t:t+\tau}\right)
\,\middle|\,
s
\right],
\label{eq:skill_distribution}
\end{equation}
where $\phi(\cdot)$ extracts spectral, temporal, and cross-receiver statistics
from a window of duration $\tau$.  Two skills are physically separable when
their induced distributions differ more than the nuisance variation introduced
by temperature, cycle state, receiver noise, and ambient RF:
\begin{equation}
d(\mathcal{P}_s,\mathcal{P}_{s'})
>
d_{\mathrm{drift}} + d_{\mathrm{noise}} .
\label{eq:separability}
\end{equation}
Equation~\ref{eq:separability} is not assumed to hold for all skill pairs.  It
is an empirical condition that motivates both our band survey and our staged
detector design.

\subsection{Workflow Hijacking as Skill-Sequence Deviation}

The security object in an LLM-agent system is not an isolated process, but an
ordered workflow.  A benign workflow consists of a planner-authorized skill
sequence, while a hijacked workflow deviates from that sequence.  Recent attacks
on RAG systems, tool documentation, and agent execution show that adversaries
can induce such deviations while preserving high-level semantic plausibility:
PoisonedRAG~\cite{poisonedrag2025}, ObliInjection~\cite{oblinjection2026},
GRAGPOISON~\cite{graphragfire2026}, and ToolHijacker~\cite{toolhijacker2026}
all demonstrate ways to redirect agent behavior through poisoned context,
retrieval, or tool interfaces.

\begin{figure}[t]
  \centering
  \includegraphics[width=\linewidth]{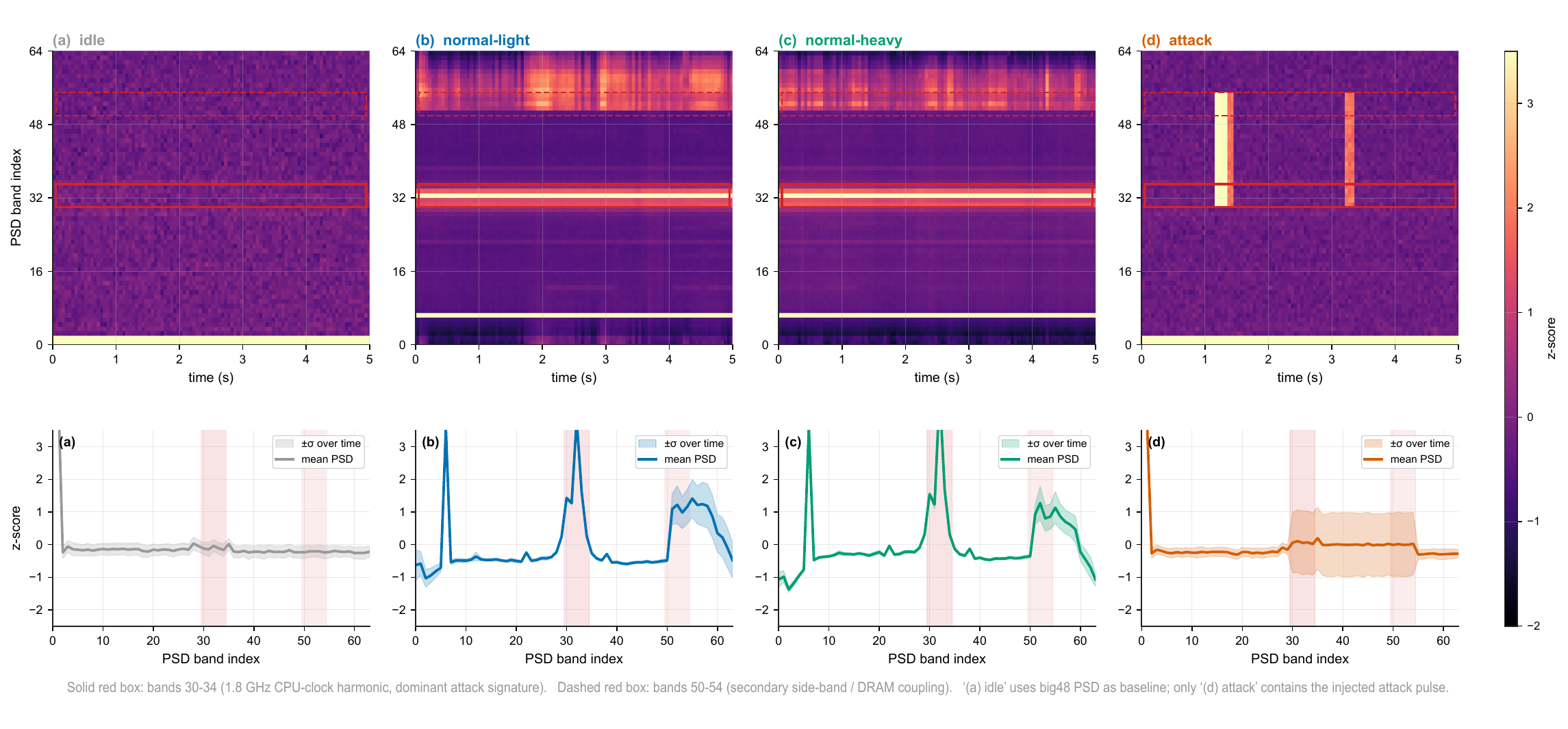}
  \caption{Power spectral density (top) and spectrogram (bottom) of two representative skills (\texttt{db\_analytics} vs.\ \texttt{log\_rotate\_compress}) on the CPU-correlated band. Skill identity manifests as both stationary spectral shape and time--frequency burst patterns.}
  \label{fig:psd-spec}
\end{figure}

\begin{figure}[t]
  \centering
  \includegraphics[width=\linewidth]{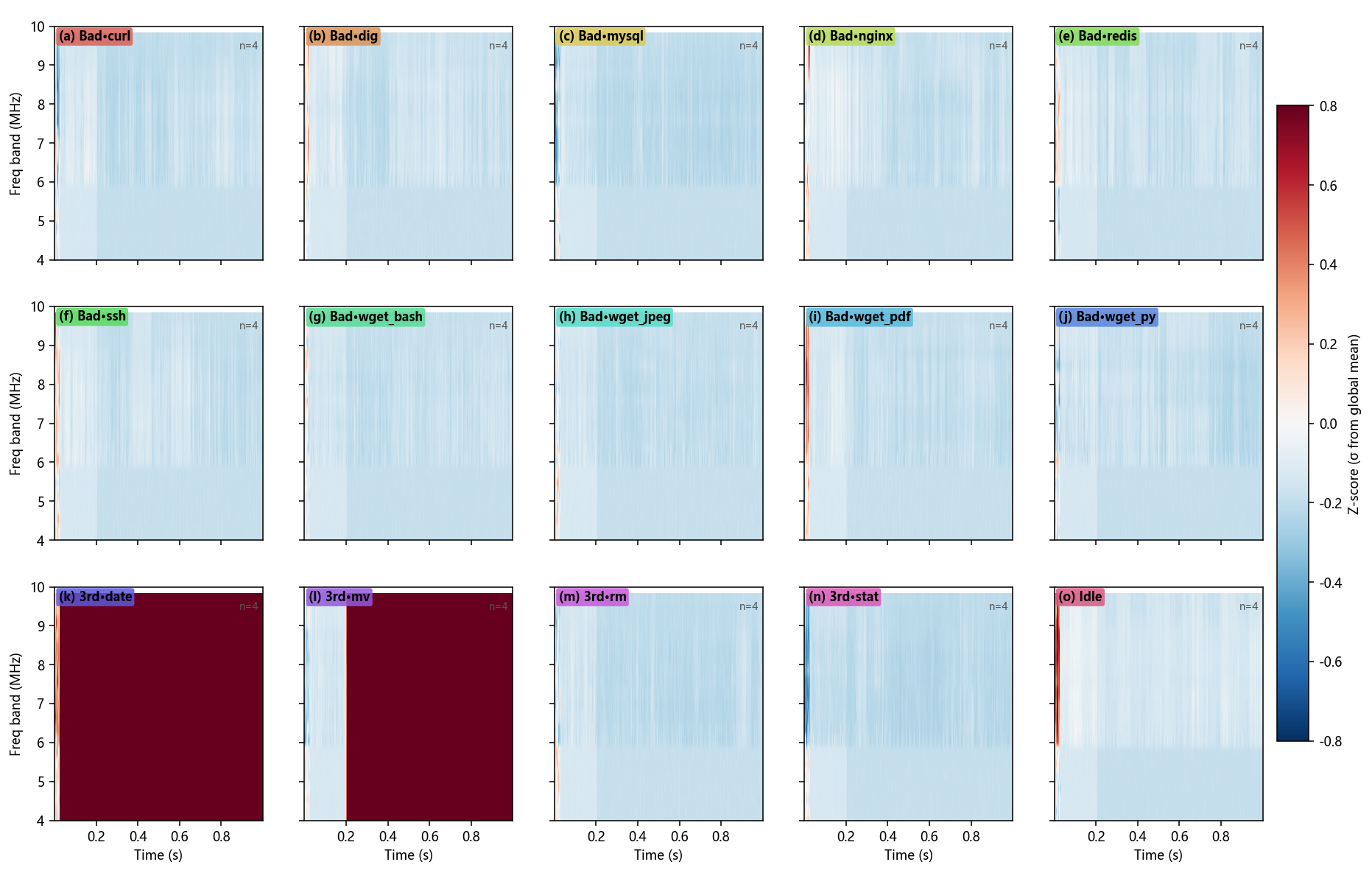}
  \caption{\new{Band-aggregated time--frequency fingerprints of the $15$-skill attack catalog.}}
  \label{fig:band-stft}
\end{figure}

We use \emph{workflow hijacking} to denote unauthorized changes to the intended
skill sequence.  The main forms are:
\begin{enumerate}[leftmargin=1.5em]
  \item \textbf{Insertion}: an unauthorized skill is added;
  \item \textbf{Omission}: a required skill, such as validation or logging, is skipped;
  \item \textbf{Substitution}: a benign skill is replaced by a different one;
  \item \textbf{Reordering}: the same skills execute in a harmful order;
  \item \textbf{Branch injection}: a covert malicious branch is added;
  \item \textbf{Parameter manipulation}: the skill identity remains but its arguments change;
  \item \textbf{Tool-result poisoning}: a legitimate tool returns corrupted output to the planner.
\end{enumerate}

From \projname's perspective, these attacks matter when they alter the physical
execution trace.  The monitor does not infer whether a natural-language plan is
morally benign.  Instead, it checks whether the hardware activity emitted by
the host is consistent with the intended skill-level workflow.  This distinction
keeps the problem measurable: semantic attacks become observable only through
their downstream effect on host execution.

\subsection{Empirical Evidence for Skill-Level EM Separability}

The above theory predicts that agent skills can be distinguishable when they
induce different mixtures of CPU activity, memory traffic, storage I/O, network
blocking, and idle intervals.  We validate this prediction in three ways.

First, empirical spectral analysis reveals that representative skills such as
\texttt{db\_analytics} and \texttt{log\_rotate\_compress} differ in both
stationary spectral shape and time-frequency burst structure, as shown in Fig.~\ref{fig:psd-spec}.  Extending this
observation to the attack-skill catalog, we note that different shell-backed and 
bad-tool-result skills produce distinct band-aggregated spectrograms.
These visual profiles support the central premise of \projname: the skill is not
merely a software label, but a workload class with a measurable physical
envelope.

Second, the informative carrier bands are deployment-specific and must be
measured rather than assumed.  Our original corpus used the
$(1.8\,\mathrm{GHz},105\,\mathrm{MHz})$ pair, motivated by the nominal CPU-clock
harmonic and a presumed DRAM-related band.  The later $1$--$3000\,\mathrm{MHz}$
survey (shown in Fig.~\ref{fig:band-stft}.) shows that the $1.8\,\mathrm{GHz}$ channel was
partly a governor-driven activity-frequency-modulation artifact, while the
$(80\,\mathrm{MHz},800\,\mathrm{MHz})$ pair gives a more defensible CPU/DRAM
separation under fixed-frequency methodology.  Figure~\ref{fig:scatter}
summarizes this carrier-selection result.  This is why \projname\ treats band
selection as a calibration step for each rig.

Third, the limitations bound the theory.  Our stress testing across the 
$16$-class \texttt{big48} experiment and the full $22$-class campaign 
(discussed in \S\ref{sec:eval:robustness}) shows that skill-level EM recognition is not
obtained by simply training a larger classifier.  Cycle state, thermal drift,
small per-class sample sizes, and cross-run distribution shift can dominate
class information.  These failures motivate the final architecture: \projname\
uses a drift-aware coarse--fine front-end, performs skill recovery only where
the class set is physically separable, and supplements sequence comparison with
a focused fine-window attack detector.

\subsection{Out-of-Band Trust Anchor}

Host-based workflow monitors rely on software telemetry: system calls,
provenance graphs, audit logs, kernel hooks, or eBPF programs
\cite{holmes2019sp,han2020unicorn,wang2020provdetector,kobra2023ndss}.  These
mechanisms are valuable when the OS is trusted, but they share the same trust
boundary as the workload they observe.  If an attacker escalates to root or
kernel privilege, the host can forge logs, suppress events, modify binaries,
or blind in-kernel monitors.  This is the symmetric-monitoring failure that
motivates \projname.

A passive EM receiver changes the trust boundary.  It does not execute code on
the device under test, does not require host-side instrumentation, and does not
consume telemetry generated by the compromised OS.  Instead, it measures an
external physical projection of hardware activity and compares the recovered
trace with an out-of-band intended workflow.

This guarantee is deliberately narrow.  EM monitoring is independent of
host-side software, not immune to all physical attacks.  A root attacker may
change CPU-governor settings or shape workloads to reduce separability; a
physical attacker may jam the RF channel or tamper with sensors.  These
adaptive cases are discussed in \S\ref{sec:discussion}.  The threat model in
the next section therefore focuses on the setting where the host software may
be fully compromised, but the external receiver and policy channel remain
outside the attacker's control.

\section{Problem Statement}
\label{sec:problem}

\subsection{Workflow Integrity Target}

Let $\mathcal{S}$ denote the skill alphabet exposed by an agent runtime.  An
intended workflow is an ordered sequence
\begin{equation}
W = \langle s_1, s_2, \ldots, s_n \rangle, \qquad s_i \in \mathcal{S},
\end{equation}
obtained from a trusted out-of-band policy channel.  The host may instead
execute a physical workflow
\begin{equation}
W^\star = \langle s^\star_1, s^\star_2, \ldots, s^\star_m \rangle,
\end{equation}
where $m$ may differ from $n$ due to insertion, omission, branch injection, or
other workflow-hijacking behavior.

The defender does not trust host-side telemetry to reveal $W^\star$.  Instead,
the defender observes passive EM traces
\begin{equation}
X = \{X_r(t; f_r)\}_{r=1}^{R},
\end{equation}
captured by external receivers.  The goal is to decide whether the physical
execution represented by $X$ is consistent with the intended workflow $W$.

\subsection{Physical Trace Recovery}

The first task is to map noisy EM observations to skill-level evidence.  Given
a windowed feature sequence
\begin{equation}
Z = \langle z_1, z_2, \ldots, z_q \rangle,
\qquad z_j = \phi(X_{t_j:t_j+\tau}),
\end{equation}
the monitor estimates either a skill label or an attack-related state for each
window.  In the ideal sequence-level setting, these predictions can be
aggregated into an estimated physical trace
\begin{equation}
\hat{W} = \langle \hat{s}_1, \hat{s}_2, \ldots, \hat{s}_{\hat{m}} \rangle .
\end{equation}

In the present system, we instantiate this abstraction in two complementary
ways.  Stage~1 performs skill recovery on physically separable skill subsets.
Stage~2 performs fine-window attack detection for short malicious sub-skills
embedded within longer task envelopes.  This design reflects the empirical
finding that large open-set multi-class skill recognition is unstable under
cycle drift and limited per-class samples.

\subsection{Integrity Decision}

At the sequence level, workflow integrity can be expressed as a comparison
between $W$ and $\hat{W}$:
\begin{equation}
g(W,\hat{W})
=
\begin{cases}
\textsc{benign}, & \mathcal{D}(W,\hat{W}) \leq \delta, \\
\textsc{hijacked}, & \mathcal{D}(W,\hat{W}) > \delta,
\end{cases}
\end{equation}
where $\mathcal{D}$ may be instantiated as a weighted edit distance over skill
insertions, deletions, substitutions, and reorderings.

However, the quantitative evaluation in this paper focuses on the physical
trace-recovery components that are implemented and measured: skill-level
classification on separable subsets and fine-window attack detection.  The
full sequence-level edit-distance comparator is part of the system design, but
its large-scale evaluation over a complete workflow attack catalog is left to
future artifact expansion.

\subsection{Scope of Detectability}

\projname\ detects workflow deviations that induce measurable changes in the
host's physical execution envelope.  It is not a semantic oracle: if a malicious
payload is deliberately shaped to be EM-indistinguishable from an authorized
skill under the selected receiver bands, passive RF monitoring alone may not
separate the two.  The security value of \projname\ is therefore an independent
physical consistency check that remains outside the compromised host's software
trust boundary.

\section{Threat Model}
\label{sec:threat_model}

We consider an enterprise environment where an LLM agent autonomously executes complex tool workflows on a host computer. The fundamental motivation for our threat model is the inherent fragility of symmetric software defenses: once the host is compromised, the very telemetry these defenses rely upon can be fundamentally manipulated from the same vantage point.

\textbf{Attacker Goals and Entry Vectors.} We assume a sophisticated adversary whose primary objective is \emph{workflow hijacking}---the unauthorized insertion, reordering, substitution, or omission of agent skills to achieve malicious ends while maintaining high-level semantic plausibility. The adversary gains initial arbitrary code execution (ACE) on the host via application-layer vectors, such as indirect prompt injection, poisoned RAG databases, or supply-chain compromises within the agent's toolchain.

\textbf{Post-Exploitation Capabilities.} Upon initial compromise, we assume the attacker successfully escalates privileges to the administrative (root) or kernel level. Within this compromised boundary, the adversary exercises full control over the host operating system. Critically, the attacker can actively \emph{blind} symmetric defenses by manipulating system binaries, intercepting system calls, bypassing in-kernel eBPF monitors, and forging or suppressing audit logs. This ensures the workflow hijacking remains completely invisible to host-internal telemetry.

\textbf{Defender Setup and Trust Anchor.} To break this symmetric trust boundary, \projname\ relocates the monitoring channel entirely out-of-band. The defender deploys a passive software-defined radio (SDR) in close, non-intrusive proximity (e.g., $2\,\mathrm{cm}$) to the target chassis. This monitor relies solely on unintentional electromagnetic (EM) emanations from the CPU and memory bus, requiring absolutely no host-side software, OS instrumentation, or physical hardware modification. 

\textbf{Out-of-Band Policy Knowledge.} We assume the defender possesses a trusted, out-of-band record of the agent's \emph{intended} workflow prior to execution. This expectation, derived from the LLM planner's original task description, is transmitted via a secure side-channel (e.g., a network-isolated policy store) and serves as the immutable ground-truth policy against which the physical EM execution trace is validated.

\textbf{Environmental Scope and Out-of-Scope Threats.} The physical deployment must contend with realistic operational conditions, including ambient electromagnetic interference, dynamic voltage and frequency scaling (DVFS) policies, and natural thermal drift. While the adversary possesses absolute software control, physical-layer attacks are strictly out of scope. We assume the attacker cannot physically alter the host's hardware, access the defender's SDR sensors, or deploy external RF transmitters to actively jam the environment.

\section{\projname\ Design}
\label{sec:design}

\subsection{Design Overview}

\projname\ is a passive RF monitor that turns electromagnetic emanations into
workflow-integrity evidence for LLM-agent executions.  Its input is a
policy-approved workflow envelope and the raw IQ streams captured by external
software-defined radios (SDRs).  Its output is a record-level verdict indicating
whether the observed physical execution is consistent with benign skill
execution or contains attack-like activity.

\begin{figure*}[t]
  \centering
  \includegraphics[width=0.8\linewidth]{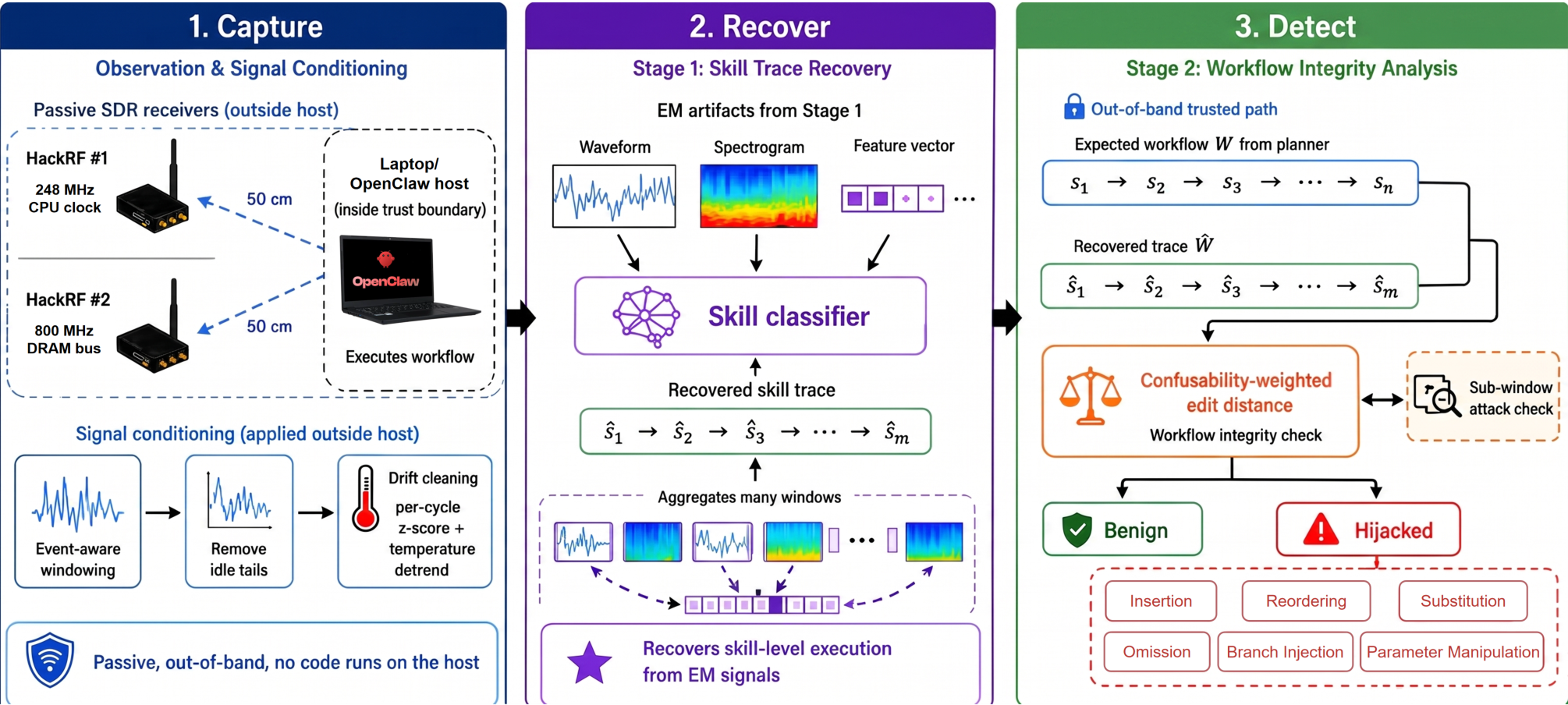}
  \caption{Overview of \projname.  Passive dual-band SDR capture feeds a
  coarse--fine windowing front-end.  Each fine window is transformed into a
  $320$-d physical feature vector, compensated for drift, and classified into
  skill-level or attack-state evidence.  Coarse-window aggregation produces the
  record-level workflow-integrity verdict evaluated in \S\ref{sec:eval}.}
  \label{fig:method_overview}
\end{figure*}

The design is intentionally staged.  Raw EM streams are noisy, continuous, and
strongly affected by temperature, cycle state, and carrier selection.  Directly
classifying an entire RF recording into a workflow verdict would mix four
separate problems: sensing, temporal alignment, drift compensation, and security
decision.  \projname\ therefore decomposes the task into the pipeline shown in
Figure~\ref{fig:method_overview}:
\begin{equation}
X
\;\longrightarrow\;
\{F_{i,j}\}
\;\longrightarrow\;
\{\mathbf{x}_{i,j}\}
\;\longrightarrow\;
\{p_{i,j}, a_{i,j}\}
\;\longrightarrow\;
v .
\label{eq:design_pipeline}
\end{equation}
Here, $X$ is the multi-receiver EM trace, $F_{i,j}$ is a fine window inside the
$i$-th skill envelope, $\mathbf{x}_{i,j}$ is its physical feature vector,
$p_{i,j}$ is skill-level evidence, $a_{i,j}$ is attack-state evidence, and
$v\in\{\textsc{benign},\textsc{hijacked}\}$ is the final verdict.

This pipeline corresponds directly to the claims evaluated in
\S\ref{sec:eval}.  Stage~1 asks whether skill-level EM envelopes are separable
on controlled skill subsets.  Stage~2 asks whether short malicious activities
embedded in longer task envelopes can be detected from fine-window RF evidence.
The production operating curve in \S\ref{sec:eval:detection} is therefore a
record-level attack-detection result, not a claim that \projname\ has fully
evaluated arbitrary sequence-level workflow verification.  This distinction is
important: the current system provides a physical consistency check for
workflow hijacking, instantiated through skill recovery and fine-window attack
detection.

The design is guided by three constraints.  First, the monitor must remain
outside the host software trust boundary: it cannot rely on system calls, audit
logs, eBPF events, or runtime self-reports.  Second, the monitored unit must be
the agent skill rather than a bit-level cryptographic operation or a whole
application.  Empirical profiling motivates this choice: different skills produce 
different macroscopic time-frequency envelopes.  Third, the inference pipeline 
must be drift-aware, because \S\ref{sec:eval:robustness} shows that cycle and 
thermal variation can dominate class information if left untreated.

\subsection{Passive RF Sensing and Carrier Calibration}
\label{sec:design:acquisition}

\projname\ observes the host through two passive SDR receivers.  The intended
role of the two channels is complementary: one channel should emphasize
CPU-correlated activity, while the other should emphasize memory-correlated
activity.  This dual view is necessary because agent skills differ not only in
average compute load, but also in their mixture of computation, DRAM traffic,
storage bursts, network blocking, and idle gaps.

Carrier choice is treated as a measured calibration problem rather than a
hardware-specification assumption.  For a candidate carrier $f$, we estimate
the workload-induced deltas
\begin{equation}
\Delta_{\mathrm{CPU}}(f)=P_{\mathrm{cpu}}(f)-P_{\mathrm{idle}}(f),
\qquad
\Delta_{\mathrm{RAM}}(f)=P_{\mathrm{ram}}(f)-P_{\mathrm{idle}}(f),
\label{eq:carrier_delta}
\end{equation}
where $P_{\mathrm{idle}}$, $P_{\mathrm{cpu}}$, and $P_{\mathrm{ram}}$ are
median band powers under idle, CPU-intensive, and memory-intensive calibration
workloads.  A useful CPU channel should have high $\Delta_{\mathrm{CPU}}$; a
useful memory channel should have high $\Delta_{\mathrm{RAM}}$ with limited CPU
confounding.

This calibration step explains the relationship between the original corpus
and the new-bands replication.  The original experiments used
$(1.8\,\mathrm{GHz},105\,\mathrm{MHz})$, motivated by the CPU-clock harmonic
and a presumed memory-related band.  The later $1$--$3000\,\mathrm{MHz}$ survey
in \S\ref{sec:eval:physical} shows that the $1.8\,\mathrm{GHz}$ channel partly
captures governor-driven activity-frequency modulation.  The measured
$(80\,\mathrm{MHz},800\,\mathrm{MHz})$ pair, summarized in
Figure~\ref{fig:scatter}, provides a more defensible CPU/DRAM separation under
fixed-frequency methodology.  The rest of \projname's pipeline is independent
of the specific carrier pair; the carrier pair is an instantiation parameter.

In the evaluated prototype, both SDRs sample at $20\,\mathrm{MS/s}$ with
$8$-bit IQ.  The original corpus uses the legacy
$(1.8\,\mathrm{GHz},105\,\mathrm{MHz})$ pair, while the replication corpus uses
the survey-selected $(80\,\mathrm{MHz},800\,\mathrm{MHz})$ pair.  Temperature is
sampled in parallel and used only for drift compensation, not as a security
label.

\subsection{Coarse--Fine Physical Evidence Extraction}
\label{sec:design:windows}
\label{sec:design:features}

The central representation in \projname\ is not a whole RF recording, but a set
of fine-window physical evidence vectors aligned to a coarse skill envelope.
A coarse window captures the expected wall-clock duration of one skill
invocation.  Inside it, \projname\ extracts overlapping fine windows:
\begin{equation}
F_{i,j}=[a_i+j\rho,\;a_i+j\rho+\tau],
\label{eq:coarse_fine_window}
\end{equation}
where $[a_i,b_i]$ is the coarse envelope of skill $i$, $\tau$ is the fine-window
length, and $\rho$ is the stride.

\begin{figure}[t]
  \centering
  \includegraphics[width=\linewidth]{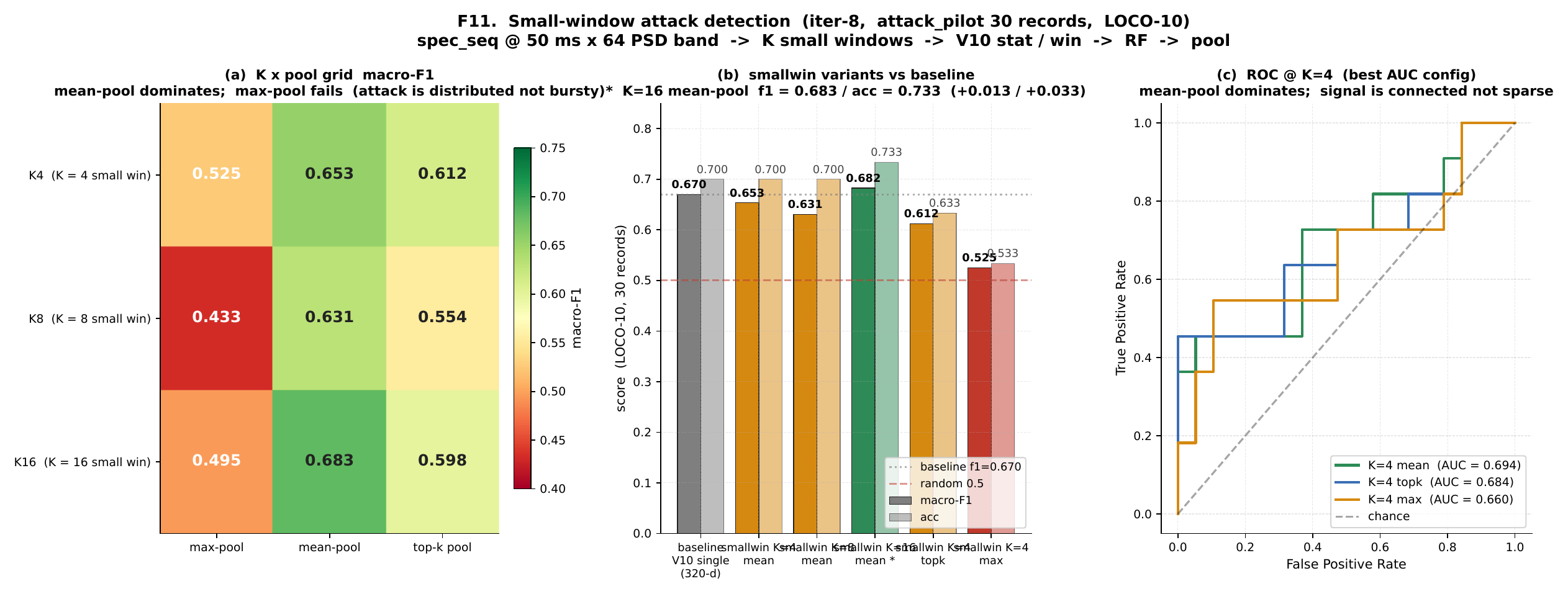}
  \caption{Coarse--fine windowing.  The coarse envelope preserves the
  skill-level task structure, while overlapping fine windows expose short
  attack payloads that would be diluted by whole-record aggregation.}
  \label{fig:smallwin}
\end{figure}

This design is motivated by the attack structure.  A malicious payload may
occupy only a short interval inside a much longer task envelope.  Whole-record
features dilute such activity, while very short global sliding windows lose
workflow context.  Coarse--fine windowing preserves both: the coarse envelope
keeps the skill-level structure, and fine windows expose localized attack
bursts.  Figure~\ref{fig:smallwin} illustrates this mechanism.

Each fine window is transformed into a $320$-dimensional V10 feature vector.
The feature groups are deliberately statistical rather than instruction-level:
\begin{itemize}[leftmargin=1.4em]
  \item \textbf{Spectral energy:} log-PSD band energies that capture stationary
  workload power differences;
  \item \textbf{Spectral shape:} centroid, bandwidth, flatness, rolloff, and
  spectral moments that capture carrier-local structure;
  \item \textbf{Temporal envelope:} amplitude moments, short-lag
  autocorrelation, and burst statistics that capture the timing of computation
  and I/O;
  \item \textbf{Cross-receiver coupling:} correlation and phase-derived
  statistics between the two SDR channels that capture CPU/DRAM co-variation.
\end{itemize}

Let $\mathbf{x}_{i,j}\in\mathbb{R}^{320}$ be the raw feature vector extracted
from $F_{i,j}$.  Before inference, \projname\ applies three leakage-controlled
preprocessing steps.  First, cycle-local normalization removes coarse amplitude
offsets:
\begin{equation}
\tilde{x}_{c,k}
=
\frac{x_{c,k}-\mu_{c,k}}{\sigma_{c,k}+\epsilon}.
\label{eq:cycle_normalization}
\end{equation}
Second, temperature detrending removes low-order thermal bias:
\begin{equation}
x'_{k}
=
\tilde{x}_{k}
-
\sum_{\ell=0}^{d}\beta_{\ell,k}T^\ell .
\label{eq:temperature_detrend}
\end{equation}
Third, ANOVA feature selection keeps the top-$k$ discriminative dimensions on
the training fold only.  The evaluated configurations use $d=1$, $k=65$ for
the focused three-skill task, and $k=80$ for larger settings.  These parameters
match the evaluation protocol in \S\ref{sec:eval:prototype}.

For supervised training and offline evaluation, the OpenClaw rig provides
event metadata that identifies the active \texttt{WORK} interval inside each
coarse record.  These events are used as labels and alignment metadata for the
physical corpus.  They are not treated as trusted runtime telemetry.  This
choice keeps the evaluated claim precise: the prototype measures whether
properly localized skill and attack intervals are physically separable in RF,
while the host-independent deployment path relies on the out-of-band workflow
schedule rather than host logs.

\subsection{Inference and Record-Level Verdict}
\label{sec:design:stage1}
\label{sec:design:stage2}

\projname\ uses the same drift-compensated feature representation for two
measured inference tasks.

\para{Stage 1: skill evidence.}
For skill-recovery experiments, a classifier maps each fine-window feature
vector to a posterior distribution over a skill set $\mathcal{S}_1$:
\begin{equation}
p_{i,j}(s)
=
\Pr(s \mid \mathbf{x}'_{i,j}),
\qquad s\in\mathcal{S}_1 .
\label{eq:skill_posterior}
\end{equation}
Fine-window posteriors are pooled inside the coarse window:
\begin{equation}
\bar{p}_{i}(s)
=
\mathrm{Pool}_{j}\;p_{i,j}(s),
\qquad
\hat{s}_i
=
\arg\max_s \bar{p}_{i}(s).
\label{eq:skill_pooling}
\end{equation}
This stage corresponds to the focused3 and pairwise/big48 results in
\S\ref{sec:eval:physical} and \S\ref{sec:eval:robustness}.  Its purpose is to test
the physical premise that some agent skills have recoverable EM envelopes.  It
is not presented as a solved open-set $22$-class recognition problem; our stress
campaigns explicitly show the limits of that setting.

\para{Stage 2: attack-state evidence.}
For workflow-hijacking detection, \projname\ uses a fine-window state detector
\begin{equation}
a_{i,j}
=
h_\psi(\mathbf{x}'_{i,j})
\in
\{\texttt{background},\texttt{normal},\texttt{attack}\}.
\label{eq:attack_state_detector}
\end{equation}
A coarse record is flagged when the aggregation of its fine-window states
indicates attack-like physical activity:
\begin{equation}
v_i
=
\begin{cases}
\textsc{hijacked}, &
A(\{a_{i,j}\}_j) > \eta,\\
\textsc{benign}, &
A(\{a_{i,j}\}_j) \leq \eta,
\end{cases}
\label{eq:record_verdict}
\end{equation}
where $A(\cdot)$ is the record-level aggregation rule and $\eta$ is the
operating threshold.  In the evaluated prototype, $A$ is implemented by
majority voting or score aggregation over fine windows, yielding the
record-level accuracy, recall, ROC, and PR results reported in
\S\ref{sec:eval:detection}.

Both stages are implemented with balanced random forests in the headline
prototype.  The default model uses $500$ trees, maximum depth $15$, and minimum
leaf size $2$.  This choice is empirical rather than architectural: tree
ensembles are stable under small physical datasets, heterogeneous feature
groups, and non-Gaussian drift. 

Table~\ref{tab:design_eval_alignment} summarizes how the design components map
to the evaluation sections.  This alignment is intentional: every mechanism in
the design is either evaluated directly or identified as an implementation
parameter, avoiding unmeasured sequence-verifier claims.

\begin{table*}[t]
  \centering
  \small
  \caption{Alignment between \projname's design components and the evaluation.
  The method is written around the mechanisms that are quantitatively evaluated
  in the paper.}
  \label{tab:design_eval_alignment}
  \begin{tabularx}{\linewidth}{lXX}
    \toprule
    \textbf{Design component} & \textbf{Prototype instantiation} & \textbf{Evaluation support} \\
    \midrule
    Carrier calibration
      & Legacy $(1.8\,\mathrm{GHz},105\,\mathrm{MHz})$ and surveyed
        $(80\,\mathrm{MHz},800\,\mathrm{MHz})$
      & Band survey and new-bands replication
        (\S\ref{sec:eval:physical}, \S\ref{sec:eval:detection}) \\
    Coarse--fine windowing
      & $20\,\mathrm{s}$ coarse envelope; $0.5\,\mathrm{s}$ fine windows with
        $0.25\,\mathrm{s}$ stride in the original pipeline
      & Stage-2 record-level detection and case studies
        (\S\ref{sec:eval:detection}, \S\ref{sec:eval:robustness}) \\
    V10 physical features
      & $320$-d dual-receiver spectral, temporal, and cross-channel statistics
      & Feature attribution and drift diagnosis
        (\S\ref{sec:eval:robustness}) \\
    Drift compensation
      & Cycle-local normalization, degree-1 temperature detrending, ANOVA
        top-$k$ selection
      & LOCO protocol and ablations
        (\S\ref{sec:eval:prototype}, \S\ref{sec:eval:robustness}) \\
    Skill evidence
      & Random-forest posterior pooling over fine windows
      & focused3 and big48/pairwise skill recovery
        (\S\ref{sec:eval:physical}, \S\ref{sec:eval:robustness}) \\
    Attack-state evidence
      & Fine-window \{\texttt{background}, \texttt{normal},
        \texttt{attack}\} detector with record-level aggregation
      & Production ROC/PR and new-bands attack recall
        (\S\ref{sec:eval:detection}) \\
    \bottomrule
  \end{tabularx}
\end{table*}

\section{Evaluation}
\label{sec:eval}

We evaluate \projname\ as a complete physical monitoring system for
LLM-agent workflow integrity.  The evaluation is organized around four
claims that match the system design in \S\ref{sec:design} and the paper's
central claim in the introduction.

\begin{itemize}[leftmargin=1.4em]
  \item \textbf{C1: Implementability.} \projname\ can be deployed as a passive
  dual-SDR monitor on real agent hosts and can collect synchronized RF,
  temperature, and workflow records at scale.
  \item \textbf{C2: Physical validity.} Agent skills induce measurable
  macroscopic EM envelopes, and informative RF bands can be selected by
  measurement rather than assumed from hardware specifications.
  \item \textbf{C3: Detection efficacy.} Fine-window physical evidence can
  detect workflow-hijacking activity at the record level with low false
  positives.
  \item \textbf{C4: Robustness and practicality.} The pipeline remains useful
  under carrier re-selection, drift-aware evaluation, and realistic runtime
  constraints.
\end{itemize}

Table~\ref{tab:eval-roadmap} maps these claims to the corresponding datasets,
figures, and measurements.  This structure is deliberate: every result in the
evaluation validates one part of the implemented system, and we avoid claiming
an unevaluated arbitrary sequence-level verifier.

\begin{table*}[t]
  \centering
  \small
  \caption{Evaluation roadmap.  Each block validates a concrete system claim
  and corresponds to the abstractions introduced in \S\ref{sec:design}.}
  \label{tab:eval-roadmap}
  \begin{tabularx}{\linewidth}{lXX}
    \toprule
    \textbf{Claim} & \textbf{What is validated} & \textbf{Evidence} \\
    \midrule
    C1: Implementability
      & Passive RF prototype, synchronized corpus, real host deployment
      & Hardware stack, main $7.82\,\mathrm{TB}$ corpus, new-bands $440\,\mathrm{GB}$ session \\
    C2: Physical validity
      & Carrier selection and skill-level EM separability
      & Band survey, PSD/spectrograms, pairwise skill separability \\
    C3: Detection efficacy
      & Record-level workflow-hijacking detection
      & Production ROC/PR, coarse--fine attack detection, case studies \\
    C4: Robustness/practicality
      & Cross-carrier replication, drift analysis, calibration, latency
      & New-bands accuracy, cycle-leakage diagnosis, calibration, runtime breakdown \\
    \bottomrule
  \end{tabularx}
\end{table*}

\subsection{End-to-End Prototype and Corpus}
\label{sec:eval:prototype}

\begin{figure}[t]
  \centering
  \begin{minipage}{0.24\textwidth}
    \centering
    \includegraphics[width=\linewidth]{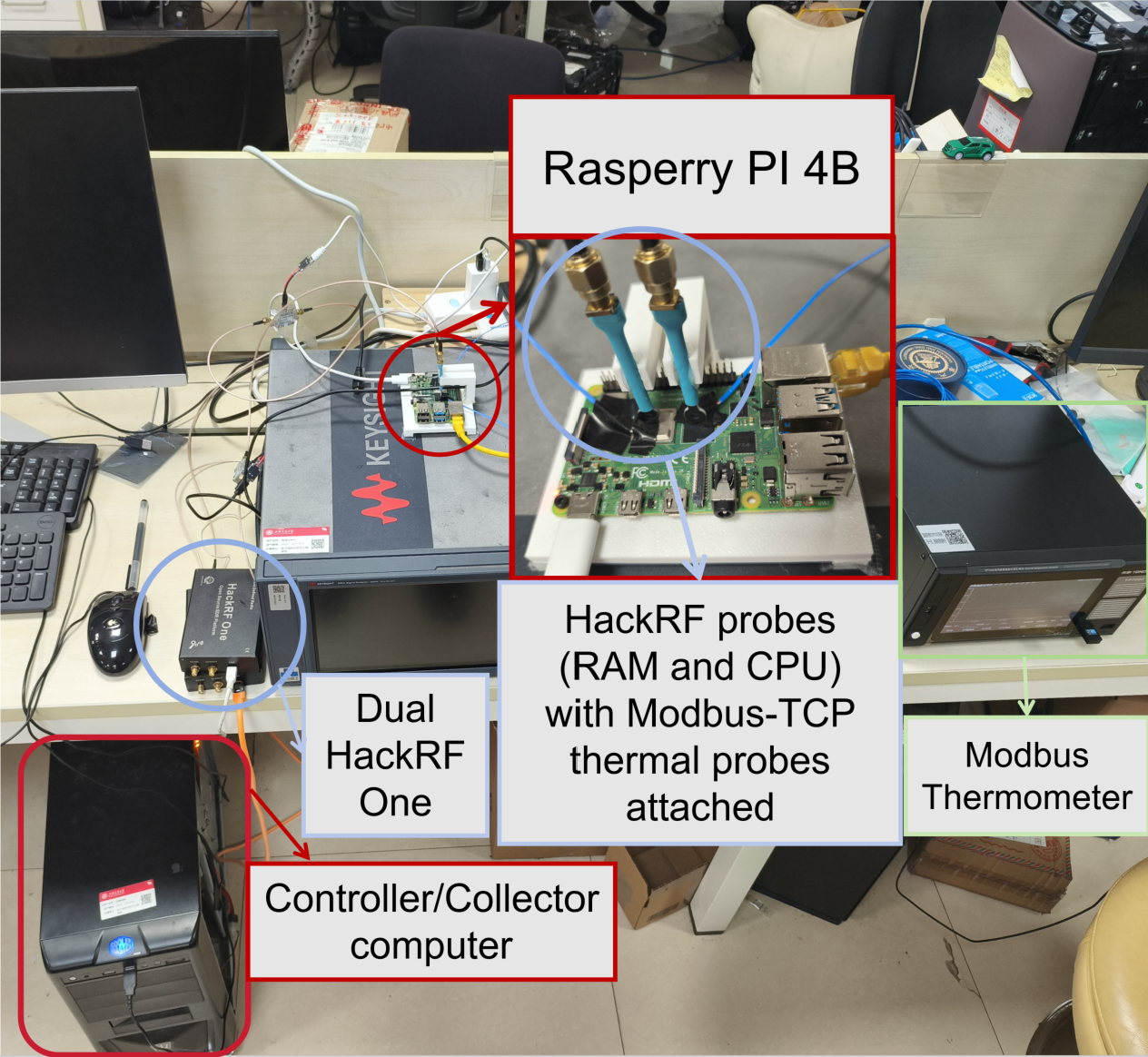}
    \caption*{Raspberry Pi target}
  \end{minipage}
  \hfill
  \begin{minipage}{0.24\textwidth}
    \centering
    \includegraphics[width=\linewidth]{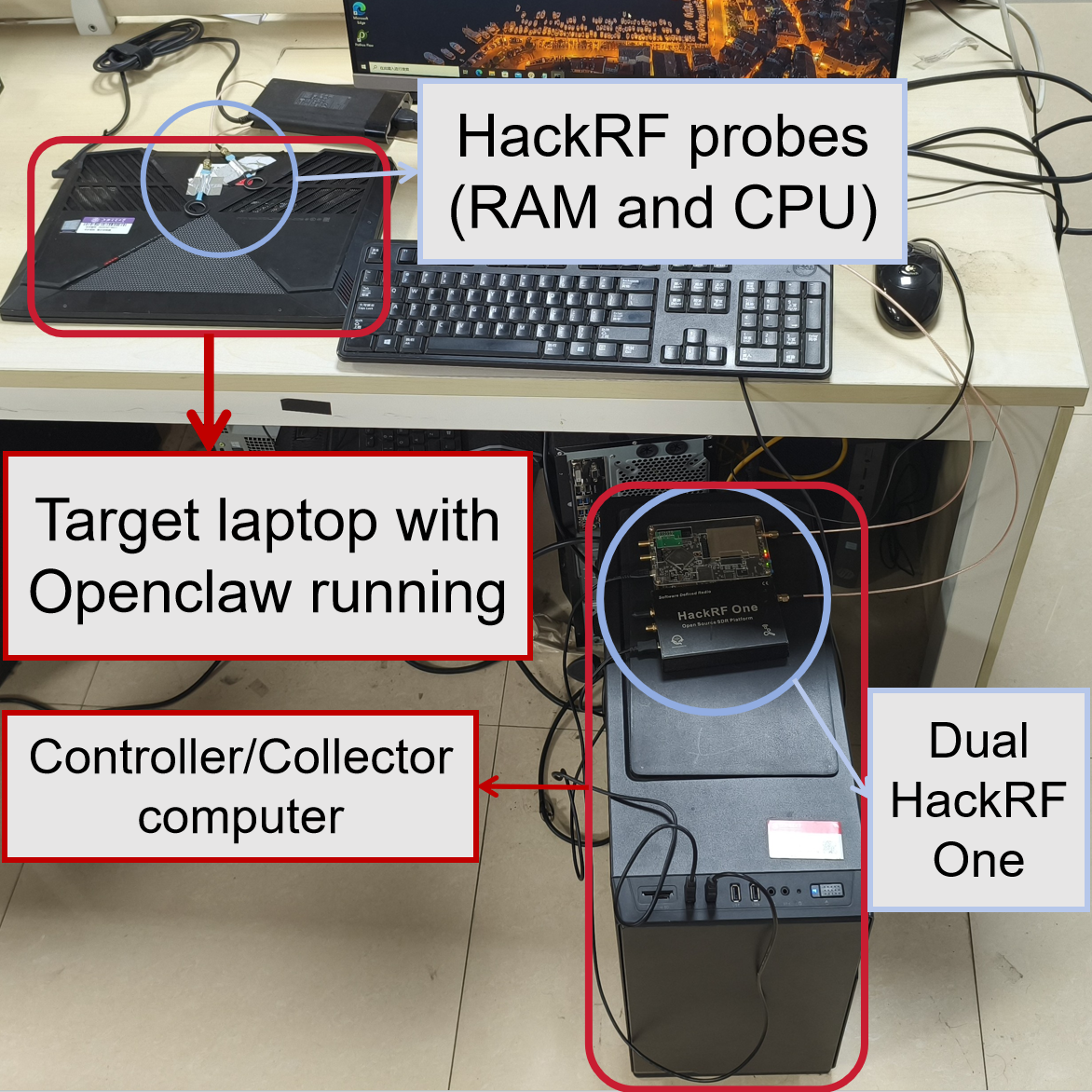}
    \caption*{Laptop target}
  \end{minipage}
  \caption{Prototype deployments.  \projname\ passively monitors real hosts
  using external SDRs.  The Laptop deployment supports the headline and
  new-bands results, heavily emphasizing realistic enterprise scenarios; the Pi
  deployment is used as an additional stress setting.}
  \label{fig:hardware-stack}
\end{figure}

\paragraph{Prototype.}
We implement \projname\ with two HackRF One SDRs, each sampling at
$20\,\mathrm{MS/s}$ with $8$-bit IQ.  The radios are placed near the target
host without any electrical or software connection to it.  A temperature sensor
records thermal state for drift compensation.  Figure~\ref{fig:hardware-stack}
shows the two physical deployments used in the evaluation: a laptop target for 
the headline and new-bands corpora, establishing it as the prominent platform, 
and a Raspberry Pi agent host for an additional portability stress campaign. 
These deployments exercise the full sensing path: passive RF capture, temperature 
logging, feature extraction, model inference, and record-level verdict generation.

\paragraph{Corpora.}
Table~\ref{tab:datasets} summarizes the RF corpora.  The main corpus contains
$12{,}232$ records and approximately $7.82\,\mathrm{TB}$ of raw IQ, covering
$16$ benign skills and $22$ attack skills.  The new-bands corpus is a separate
re-collection using the measured $(80,800)\,\mathrm{MHz}$ carrier pair.  It
runs for $98$ minutes over $10$ cycles and produces $550$ IQ files
($276$ CPU-channel and $274$ RAM-channel files), totaling $440\,\mathrm{GB}$ of
raw IQ.  The session also records $221$ per-skill event files and $11{,}114$
temperature samples.  The Pi-side \texttt{openclaw\_attack\_v1}
corpus is used only as a stress campaign because it uses a third carrier pair
and has small per-class support.

\begin{table*}[t]
  \centering
  \small
  \caption{Evaluation corpora.  The main corpus supports the headline results.
  The new-bands corpus validates carrier re-selection.  The Raspberry Pi
  campaign is used as a boundary stress test, not as a headline accuracy
  source.}
  \label{tab:datasets}
  \setlength{\tabcolsep}{7pt}
  \begin{tabular}{lccccc}
    \toprule
    Dataset & Host & Skills & Records / files & Trace & Carrier pair \\
    \midrule
    \texttt{focused3}
      & Laptop & $3$ benign & $123$ records & $8\,\mathrm{s}$
      & $1.8\,\mathrm{GHz},105\,\mathrm{MHz}$ \\
    \texttt{big48}
      & Laptop & $16$ benign & $1{,}529$ records & $8\,\mathrm{s}$
      & $1.8\,\mathrm{GHz},105\,\mathrm{MHz}$ \\
    \texttt{attack\_skills\_pilot}
      & Laptop & $22$ attacks & $299$ records & $20\,\mathrm{s}$
      & $1.8\,\mathrm{GHz},105\,\mathrm{MHz}$ \\
    Main corpus
      & Laptop & $16+22$ & $12{,}232$ records & mixed
      & $1.8\,\mathrm{GHz},105\,\mathrm{MHz}$ \\
    New-bands corpus
      & Laptop & $22$ attacks + idle & $550$ IQ files & $20\,\mathrm{s}$
      & $80\,\mathrm{MHz},800\,\mathrm{MHz}$ \\
    \texttt{openclaw\_attack\_v1}
      & Pi & $22$ attacks & $155$ records & $20\,\mathrm{s}$
      & $248\,\mathrm{MHz},800\,\mathrm{MHz}$ \\
    \bottomrule
  \end{tabular}
\end{table*}

\paragraph{Evaluation protocol.}
All reported classification results use leave-one-cycle-out (LOCO)
cross-validation unless stated otherwise.  LOCO is stricter than random
splitting because recordings in the same cycle share temperature, background
RF, and collection-order effects.  Cycle-local normalization, temperature
detrending, ANOVA feature selection, and scaling are fit only on the training
fold.  This prevents collection-time leakage from inflating results.

\subsection{Physical Validity of EM Evidence}
\label{sec:eval:physical}

This subsection validates the physical premise behind \projname: agent skills
must produce measurable EM structure, and the receiver bands must be selected
from measurements rather than assumptions.

\paragraph{Skill envelopes are visible.}
Empirical spectral analysis confirms that representative skills such as
\texttt{db\_analytics} and \texttt{log\_rotate\_compress} differ in both
stationary spectral shape and time-frequency burst structure.  Extending the observation
to the attack-skill catalog, different attack primitives produce distinct band-aggregated spectrograms.
These profiles support the paper's central physical claim that skill executions
are not merely software labels; they induce macroscopic RF envelopes.

\begin{figure}[t]
  \centering
  \includegraphics[width=\linewidth]{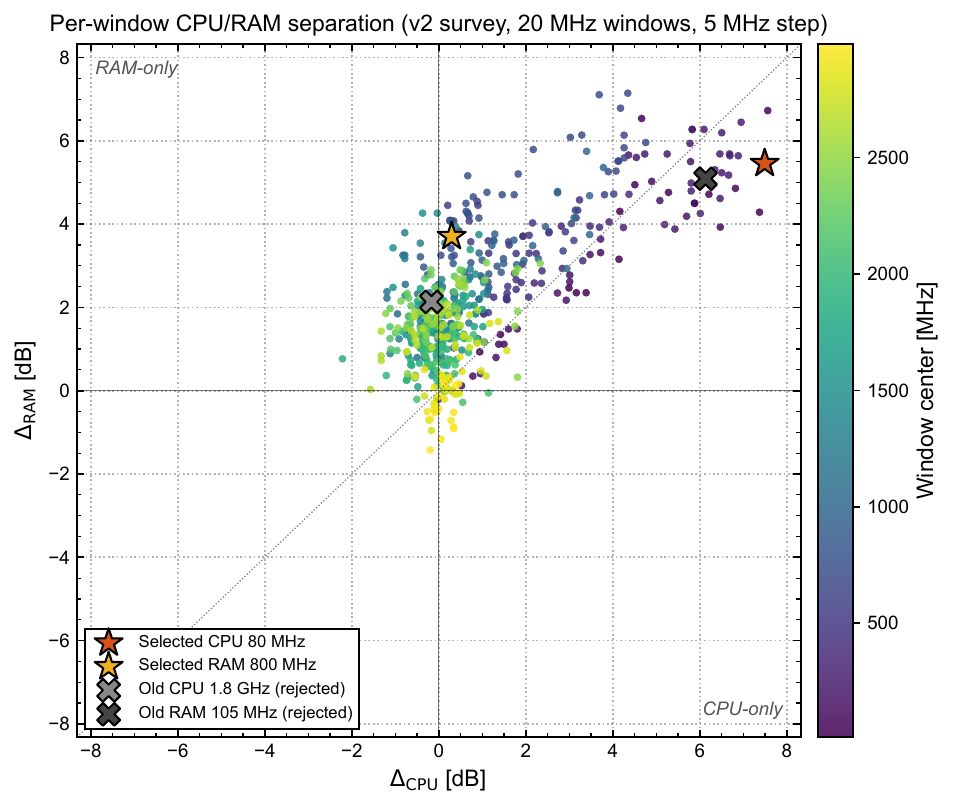}
  \caption{Carrier-separation map from the $1$--$3000\,\mathrm{MHz}$ survey.
  The selected $(80,800)\,\mathrm{MHz}$ pair provides a more defensible
  CPU/RAM split than the legacy $(1.8\,\mathrm{GHz},105\,\mathrm{MHz})$ pair.}
  \label{fig:scatter}
\end{figure}

\paragraph{Carrier selection must be measured.}
We sweep $1$--$3000\,\mathrm{MHz}$ under idle, CPU-heavy, and RAM-heavy
conditions.  For each candidate carrier $f$, we compute
\[
\Delta_{\mathrm{CPU}}(f)=P_{\mathrm{cpu}}(f)-P_{\mathrm{idle}}(f),\qquad
\Delta_{\mathrm{RAM}}(f)=P_{\mathrm{ram}}(f)-P_{\mathrm{idle}}(f).
\]
The original corpus used $(1.8\,\mathrm{GHz},105\,\mathrm{MHz})$, motivated by
the nominal CPU harmonic and a presumed memory-related band.  The survey shows
that this choice is not robust under fixed-frequency methodology: the
$1.8\,\mathrm{GHz}$ region partly captures governor-driven activity-frequency
modulation rather than a stable idle-vs-busy power signature.  The measured
$(80,800)\,\mathrm{MHz}$ pair is more defensible: $80\,\mathrm{MHz}$ is
CPU-correlated, and $800\,\mathrm{MHz}$ aligns with LPDDR4 activity.  Figure
\ref{fig:scatter} summarizes the resulting CPU/RAM separation map.

\begin{figure}[t]
  \centering
  \includegraphics[width=\linewidth]{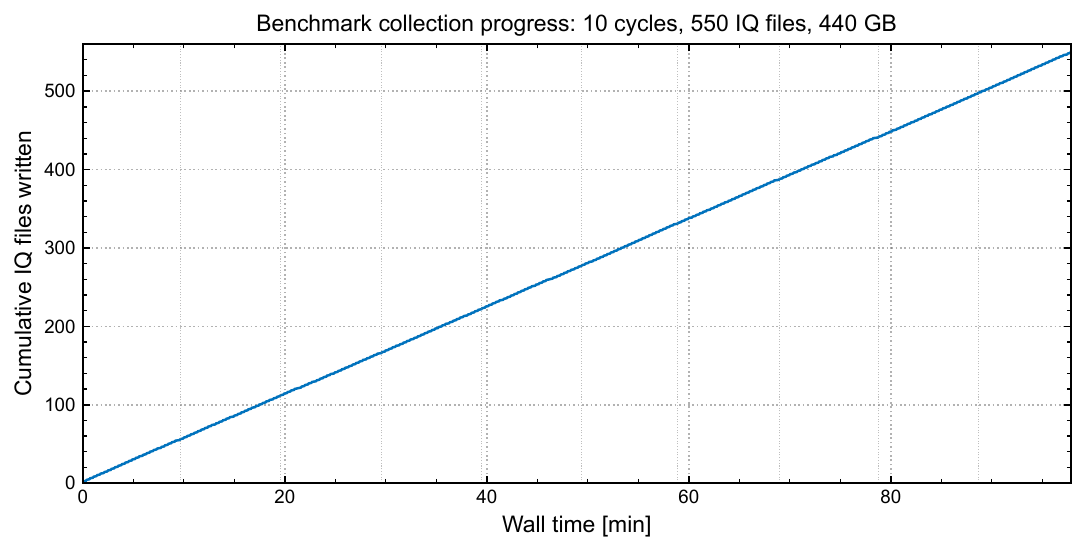}
  \caption{New-bands collection progress.  The $10$-cycle
  $(80,800)\,\mathrm{MHz}$ session runs steadily for $98$ minutes.}
  \label{fig:progress}
\end{figure}

\begin{figure}[t]
  \centering
  \includegraphics[width=\linewidth]{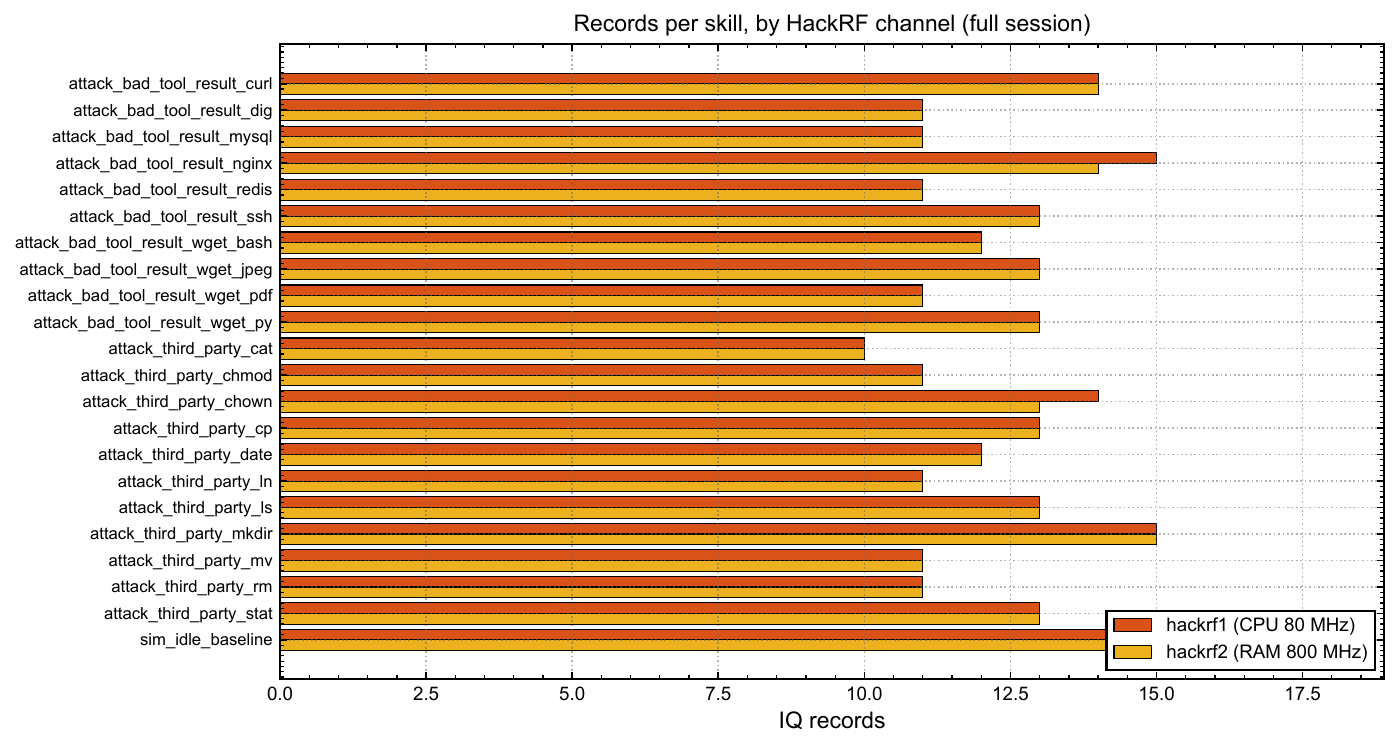}
  \caption{Per-skill record counts in the new-bands corpus.  Counts are
  balanced across SDR channels, with only minor one-record differences at
  shutdown.}
  \label{fig:per_skill_counts}
\end{figure}

\begin{figure}[t]
  \centering
  \includegraphics[width=\linewidth]{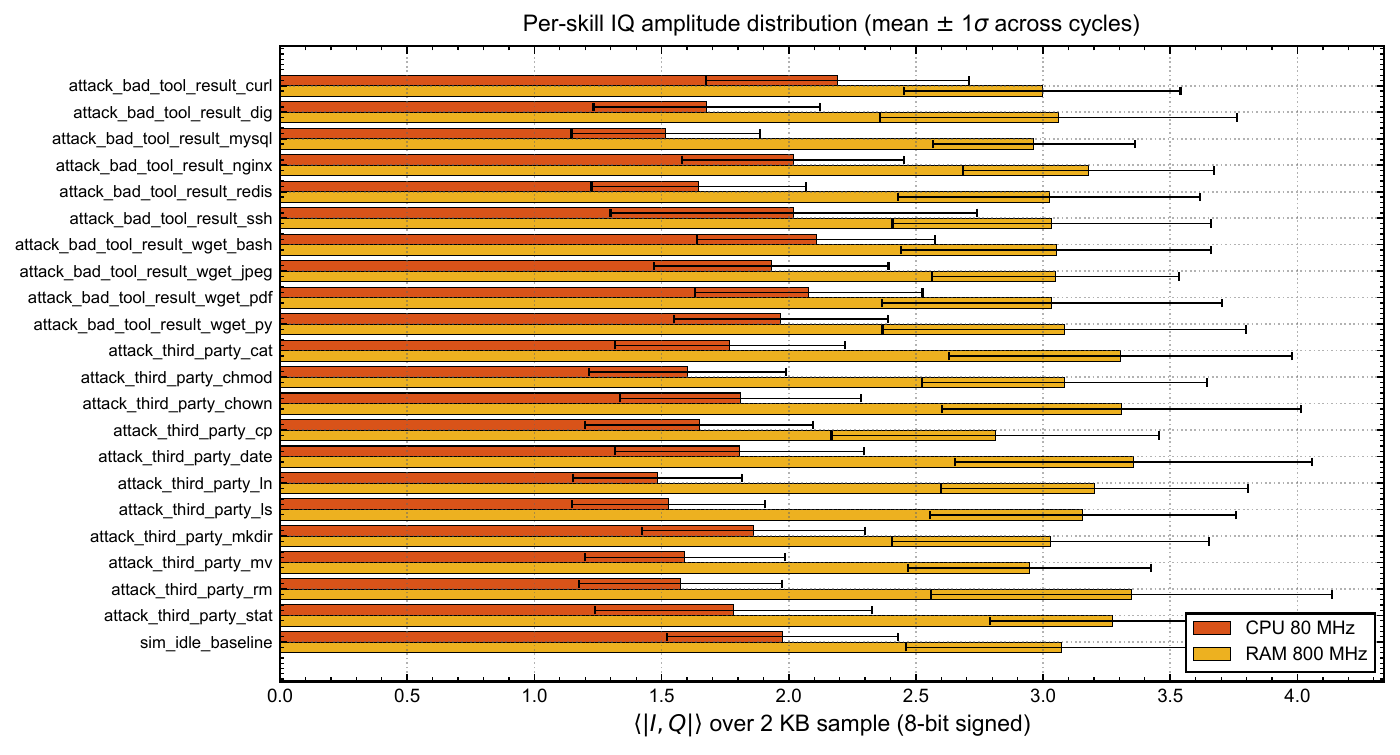}
  \caption{Per-skill IQ amplitude in the new-bands corpus.  The two receiver
  channels provide distinct and stable amplitude regimes across the session.}
  \label{fig:iq_amp}
\end{figure}

\paragraph{New-bands collection is stable.}
After selecting $(80,800)\,\mathrm{MHz}$, we re-collected a complete attack
benchmark.  Figure~\ref{fig:progress} shows steady progress across the
$98$-minute run; Figure~\ref{fig:per_skill_counts} shows balanced per-skill
record counts across channels; Figure~\ref{fig:iq_amp} shows stable per-skill
IQ amplitude.  The CPU channel sits around $1.5$--$3.0$ on the signed
$8$-bit scale, while the RAM channel sits around $3.0$--$4.5$, with small
within-skill variance for most skills.  This validates the RF collection chain
as a repeatable measurement instrument.

\begin{figure}[t]
  \centering
  \includegraphics[width=\linewidth]{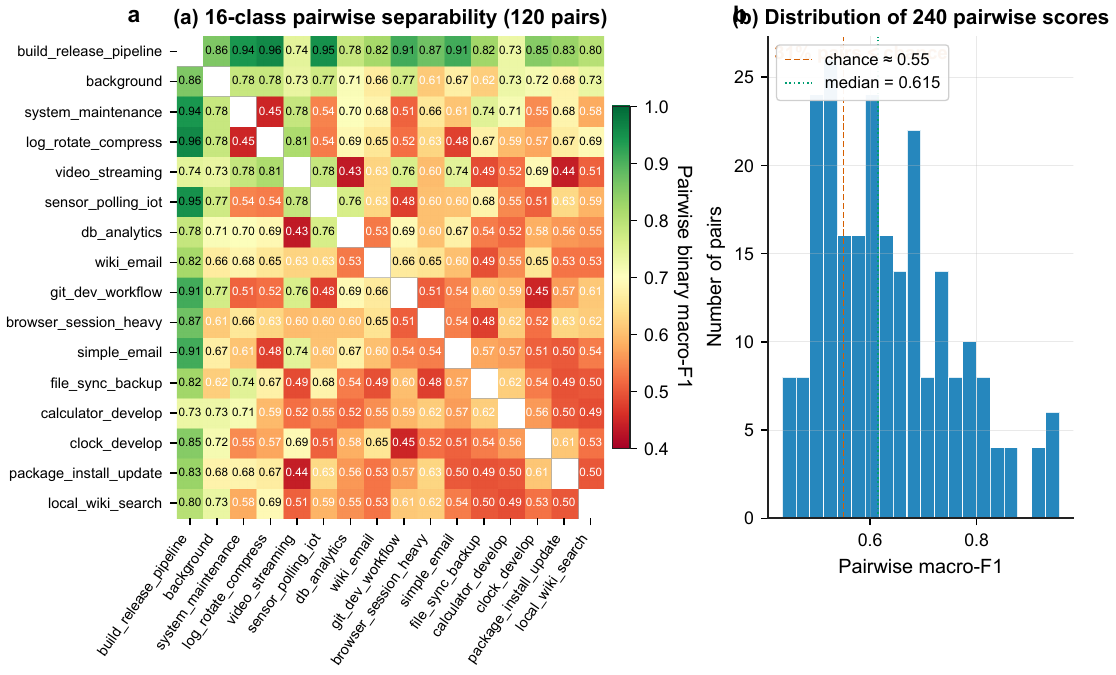}
  \caption{Pairwise skill separability on \texttt{big48}.  Skill-level EM
  evidence is structured: some pairs are highly separable, while broad
  open-set multi-class recognition remains difficult.}
  \label{fig:perskill}
\end{figure}

\paragraph{Skill separability is structured.}
On \texttt{focused3}, stage~1 reaches macro-$F_1 = 0.894 \pm 0.009$ across
five seeds under LOCO, with the worst fold above $0.878$.  This confirms that
some skill groups are recoverable from RF evidence.  The broader
\texttt{big48} catalog is more heterogeneous: a flat $16$-class classifier
reaches only macro-$F_1=0.146$.  However, pairwise analysis reveals strong
structure.  Figure~\ref{fig:perskill} shows that $12$ of $120$ pairwise
classifiers exceed $0.80$ $F_1$, and the most separable skill,
\texttt{build\_release\_pipeline}, exceeds $0.94$ against several operational
skills.  This result justifies the design choice in \S\ref{sec:design}: use
skill evidence where physical separability is strong, and use fine-window
attack evidence for localized hijacking behavior.

\subsection{Workflow-Hijacking Detection}
\label{sec:eval:detection}

This subsection evaluates the main security claim: \projname\ detects
workflow-hijacking activity from passive physical evidence.

\begin{figure}[t]
  \centering
  \includegraphics[width=\linewidth]{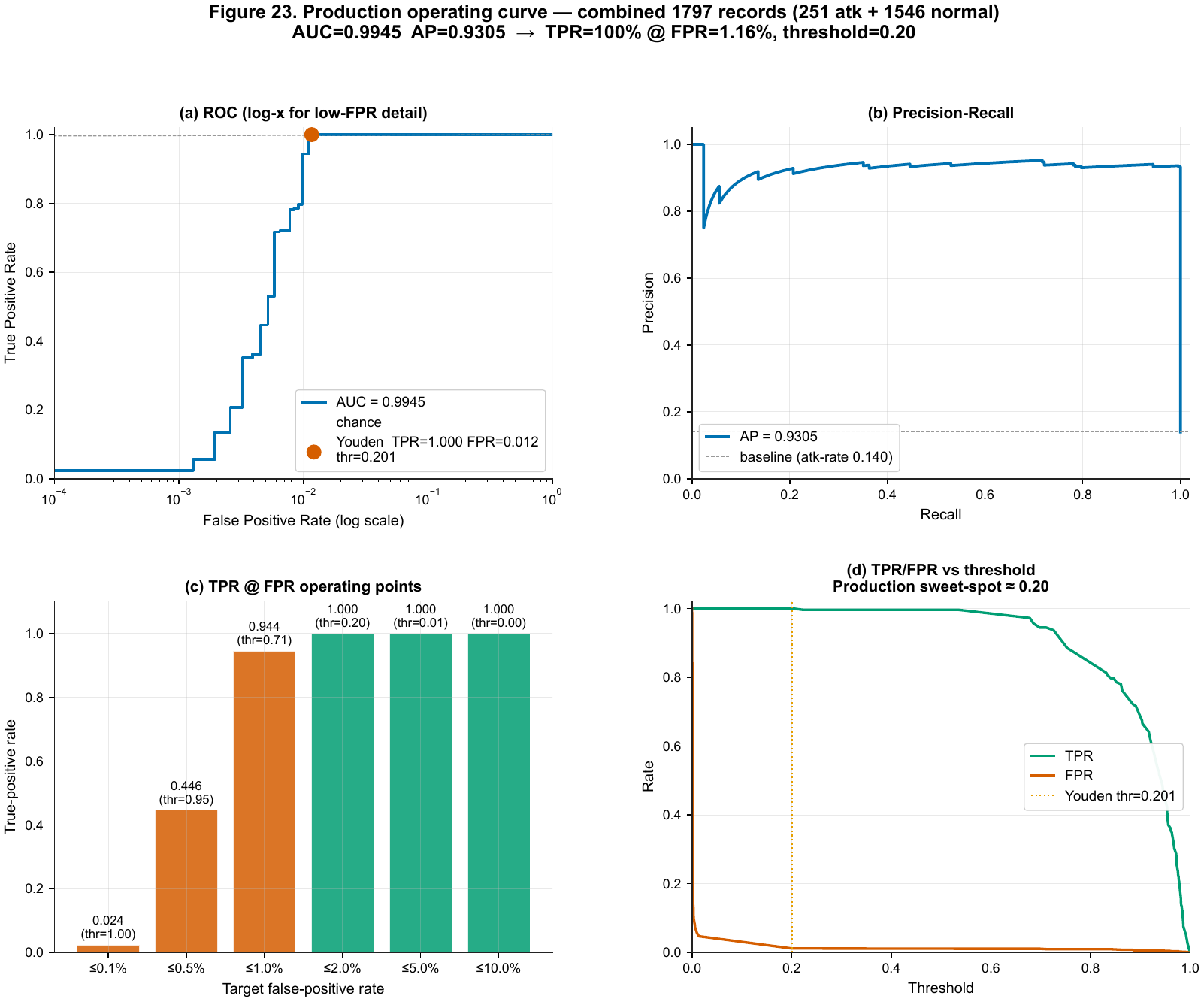}
  \caption{Production operating curve on the $11{,}800$-record split.
  \projname\ achieves $\mathrm{AUC}=0.9945$ and detects attacks at
  $100\%$ TPR with $1.16\%$ FPR.}
  \label{fig:fpr-sweep}
\end{figure}

\paragraph{Production operating curve.}
On a production split of $11{,}800$ records ($1{,}650$ attack and $10{,}150$
normal), \projname\ achieves $\mathrm{ROC\text{-}AUC}=0.9945$ and
$\mathrm{PR\text{-}AUC}=0.9305$.  At the selected operating point, it reaches
$100\%$ true-positive rate at $1.16\%$ false-positive rate.  Even under the
more conservative region $\mathrm{FPR}\leq 1\%$, TPR remains approximately
$0.99$.  Figure~\ref{fig:fpr-sweep} shows the curve.

\paragraph{Coarse--fine detection.}
Table~\ref{tab:cf} evaluates the fine-window attack detector used by the
record-level pipeline.  Across four configurations, record-level accuracy is
between $0.9252$ and $0.9398$, and attack recall is between $0.833$ and
$0.861$.  The result supports the coarse--fine design: short malicious payloads
that are diluted in a whole-record representation become visible when the
record is decomposed into fine physical windows.

\begin{table}[t]
  \centering\small
  \caption{Coarse--fine record-level attack detection on
  \texttt{big48\_chunk1 + attack\_skills\_pilot\_20260423b}.}
  \label{tab:cf}
  \begin{tabular}{lcccc}
    \toprule
    Config & Sub-win acc. & Record acc. & bg recall & atk recall \\
    \midrule
    v1 & $0.816$ & $\mathbf{0.9398}$ & $0.250$ & $0.833$ \\
    v2 & $0.787$ & $0.9252$ & $0.139$ & $0.844$ \\
    v3 & $0.809$ & $0.9398$ & $0.222$ & $0.861$ \\
    no-temp & $0.809$ & $0.9398$ & $0.222$ & $0.861$ \\
    \bottomrule
  \end{tabular}
\end{table}

\paragraph{Why anomaly detection is insufficient.}
One-class anomaly baselines do not solve the task.  IsolationForest,
OneClassSVM, and Mahalanobis-style scoring remain near chance in the combined
setting.  The reason is that ``attack'' is not equivalent to density anomaly:
benign skills are diverse, and many attack windows lie near benign workload
manifolds.  Supervised fine-window attack evidence is therefore necessary.

\paragraph{Cross-carrier detection.}
The new-bands corpus tests whether the detection pipeline survives carrier
re-selection.  On the $(80,800)\,\mathrm{MHz}$ corpus, the same V10 pipeline
reaches $83.6\%$ sub-window accuracy and $88.3\%$ record-vote accuracy.
Attack recall is $88.1\%$ at the sub-window level and $90.3\%$ at the
record level (Table~\ref{tab:newbands}).  This result is lower than the
legacy-corpus AUC, but it is more deployment-relevant because the carrier pair
is selected by measurement rather than by the governor-sensitive
$1.8\,\mathrm{GHz}$ channel.

\begin{table}[t]
  \centering\small
  \caption{LOCO performance on the new-bands $(80,800)\,\mathrm{MHz}$ corpus.
  The result is computed on the $14$ attack classes that survived event-anchor
  filtering.}
  \label{tab:newbands}
  \begin{tabular}{lcc}
    \toprule
     & Sub-window & Record vote \\
    \midrule
    Overall accuracy & $83.6\%$ & $\mathbf{88.3\%}$ \\
    Background recall & $75.0\%$ & $70.0\%$ \\
    Attack recall & $88.1\%$ & $\mathbf{90.3\%}$ \\
    \bottomrule
  \end{tabular}
\end{table}

\begin{figure}[t]
  \centering
  \begin{subfigure}[t]{0.48\linewidth}
    \centering
    \includegraphics[width=\linewidth]{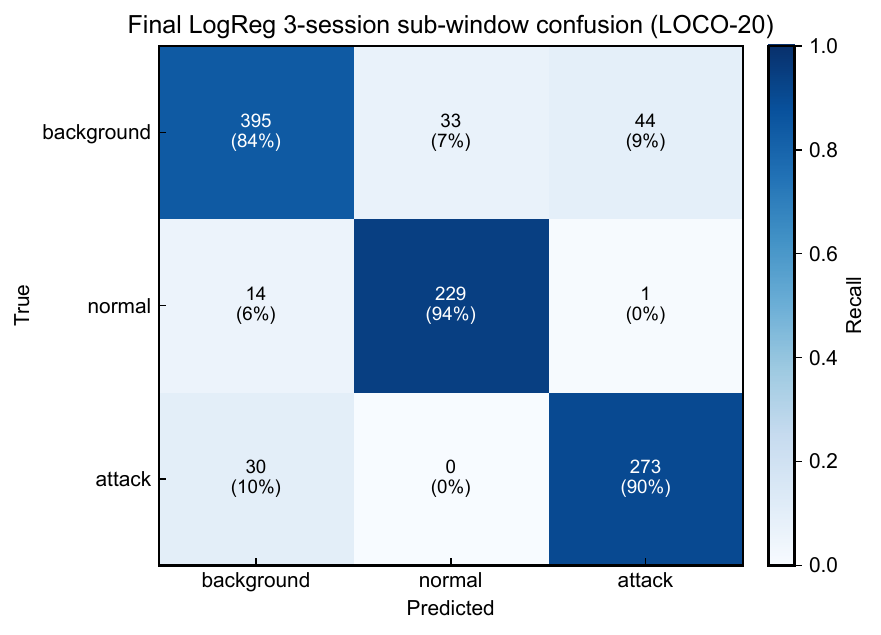}
    \caption{Sub-window}
  \end{subfigure}
  \hfill
  \begin{subfigure}[t]{0.48\linewidth}
    \centering
    \includegraphics[width=\linewidth]{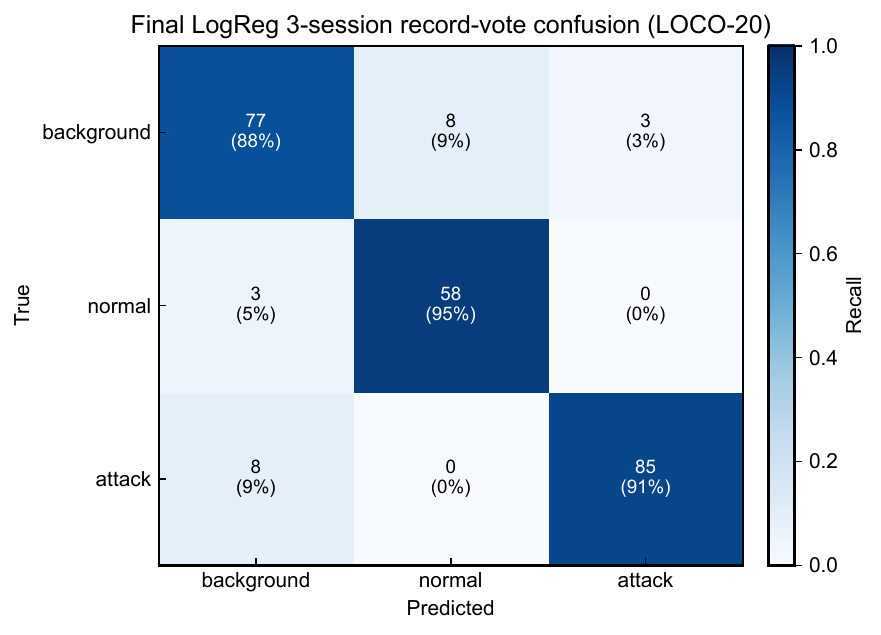}
    \caption{Record vote}
  \end{subfigure}
  \caption{New-bands confusion matrices.  Most errors are conservative:
  background windows are sometimes flagged as attack, rather than attacks being
  missed.}
  \label{fig:confusions_newbands}
\end{figure}

Figure~\ref{fig:confusions_newbands} shows that most new-bands errors are
background-as-attack, not attack-as-background.  This is the conservative
failure direction for an integrity monitor: the system is more likely to raise
a false alert than to silently miss an attack.

\subsection{Robustness, Transfer, and Cost}
\label{sec:eval:robustness}

This final subsection explains why the reported results should be read as a
measured physical system rather than a lucky split, and it states the known
boundaries of the current prototype.

\begin{figure}[t]
  \centering
  \includegraphics[width=\linewidth]{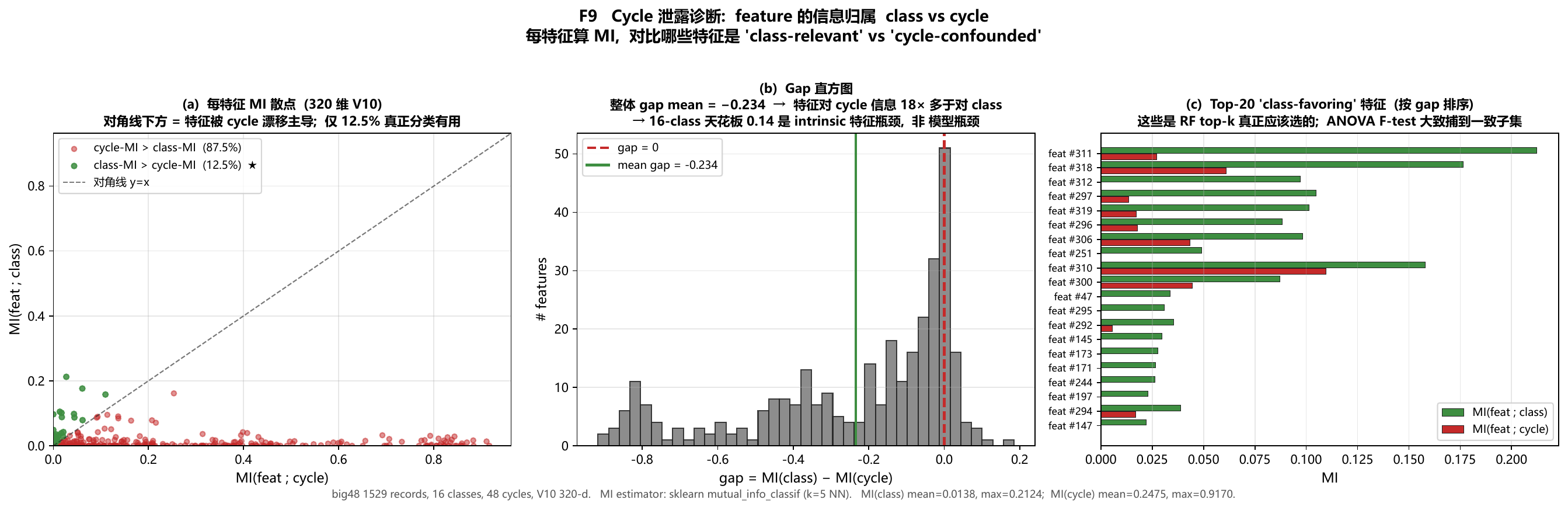}
  \caption{Cycle-leakage diagnosis.  Many RF features are more predictive of
  collection cycle than of skill identity, motivating LOCO evaluation and
  drift compensation.}
  \label{fig:cycle-leak}
\end{figure}

\paragraph{Drift is measured and controlled.}
Cycle and thermal drift are substantial.  On \texttt{big48}, the average mutual
information between V10 features and cycle index is roughly $18\times$ larger
than the mutual information with skill class.  Figure~\ref{fig:cycle-leak}
shows this effect.  This is why all headline experiments use LOCO rather than
random splits, and why the pipeline includes cycle-local normalization,
temperature detrending, and training-fold-only feature selection.

\begin{figure}[t]
  \centering
  \includegraphics[width=\linewidth]{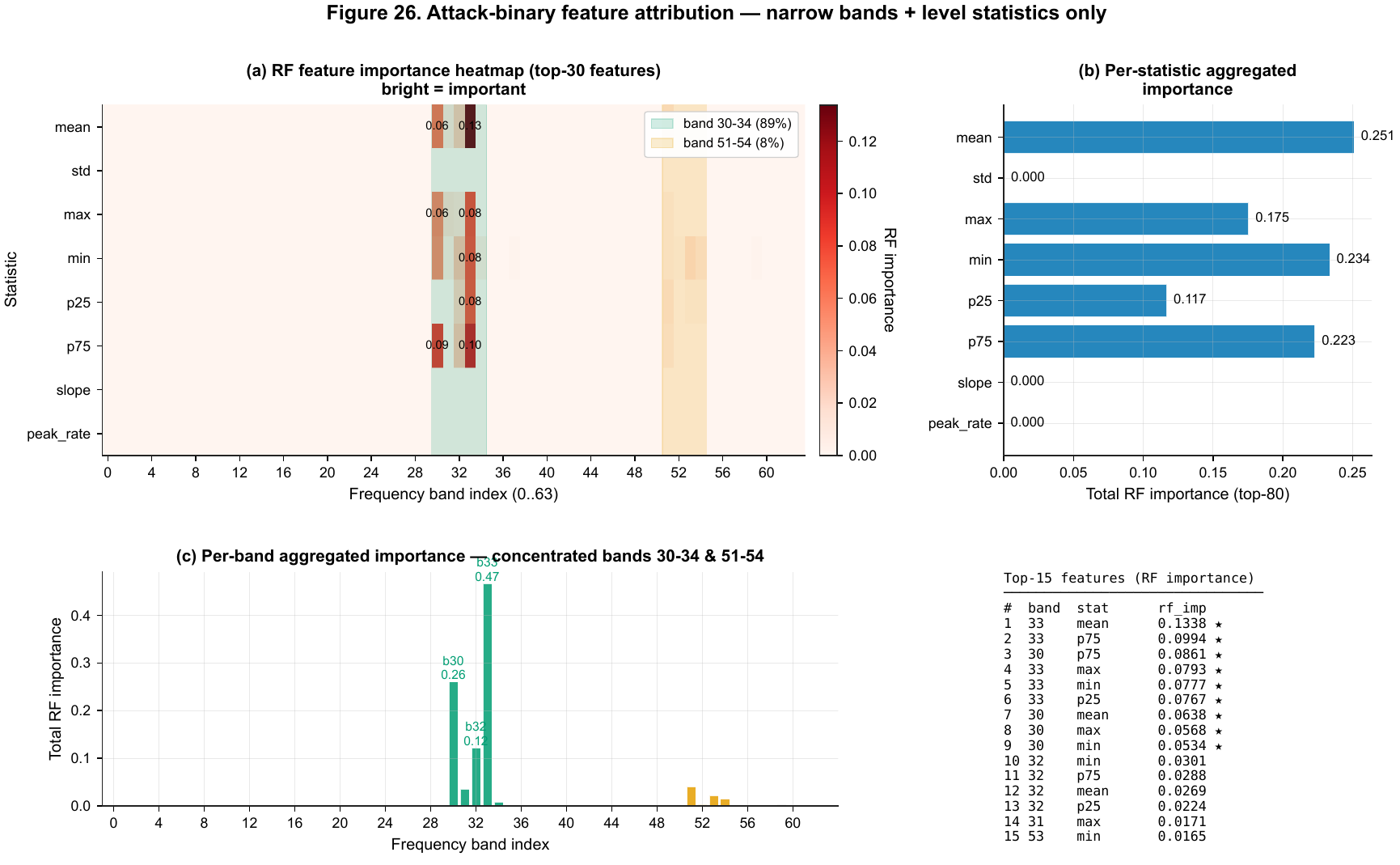}
  \caption{Band-level feature attribution on the original corpus.  The
  concentration near the $1.8\,\mathrm{GHz}$ harmonic motivates measured
  carrier re-selection and the new-bands replication.}
  \label{fig:feature-attrib}
\end{figure}

\paragraph{Feature attribution explains the legacy result.}
Figure~\ref{fig:feature-attrib} shows that the legacy model relies heavily on
bands $30$--$34$, corresponding to the $1.8\,\mathrm{GHz}$ harmonic region.
This explains the strong original-corpus AUC, but also reveals why the
new-bands replication is necessary: the legacy channel partly reflects
governor-driven activity-frequency modulation.  Spectral masking confirms that
the model is using RF structure rather than random leakage; masking the
dominant cluster degrades AUC substantially.

\begin{figure}[t]
  \centering
  \includegraphics[width=\linewidth]{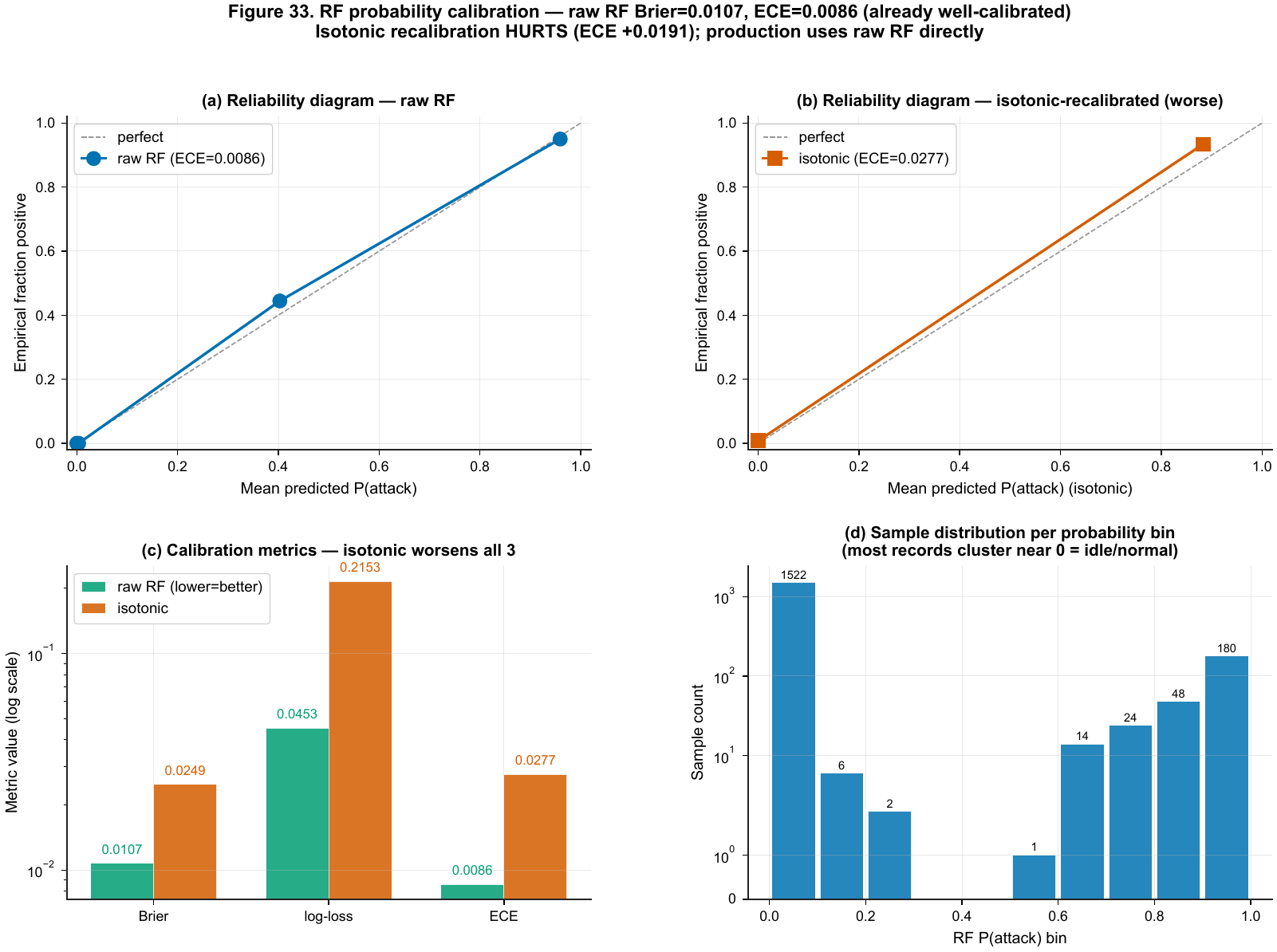}
  \caption{Probability calibration.  Raw random-forest outputs are nearly
  identity-calibrated; isotonic recalibration is unnecessary.}
  \label{fig:calibration}
\end{figure}

\paragraph{Probability outputs are stable.}
The raw random-forest probabilities are already well calibrated, with Brier
score $0.011$ and ECE $0.009$.  Post-hoc isotonic recalibration degrades the
metrics, as shown in Figure~\ref{fig:calibration}.  The production operating
point therefore uses the raw model scores.

\paragraph{Transfer stress test.}
The Pi-side \texttt{openclaw\_attack\_v1} campaign runs the
$22$-class attack catalog on a separate host and a third carrier pair
$(248,800)\,\mathrm{MHz}$.  We do not use this campaign as a headline accuracy
result because it has only $155$ records across $22$ classes and mixes host,
run, and carrier shifts.  Instead, it is a boundary test.  It shows that naive
single-shot $22$-class recognition is statistically underpowered at small
per-class sample sizes, and that cross-run transfer can collapse under drift.
This finding supports the staged design: \projname\ should not be evaluated as
a universal $22$-class classifier, but as a physical consistency monitor built
from separable skill evidence and fine-window attack evidence.

\begin{figure}[t]
  \centering
  \includegraphics[width=\linewidth]{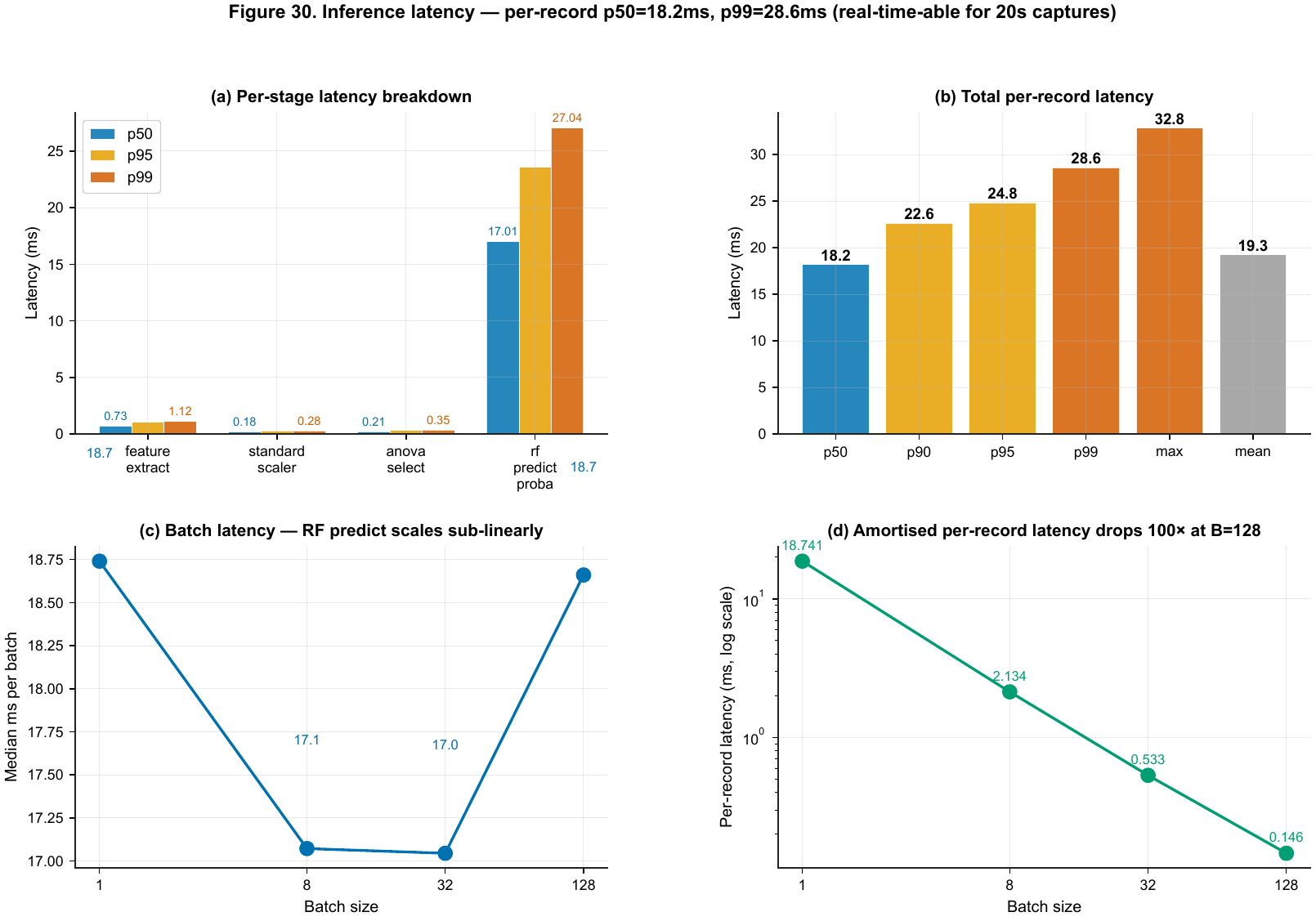}
  \caption{Inference latency.  Median post-feature latency is
  $18\,\mathrm{ms}$ and $p_{99}$ is $29\,\mathrm{ms}$.}
  \label{fig:latency}
\end{figure}

\paragraph{Runtime.}
\projname's post-feature inference latency is small relative to skill duration.
Median per-record inference latency is $18\,\mathrm{ms}$, and $p_{99}$ is
$29\,\mathrm{ms}$ (Figure~\ref{fig:latency}).  Batched prediction amortizes to
approximately $0.15\,\mathrm{ms}$ per record.  Since agent skills are
seconds-scale and fine windows are hundreds of milliseconds long, sensing and
windowing dominate wall-clock delay; model inference is not the bottleneck.

\paragraph{Evaluation takeaways.}
The evaluation supports the following claims.  First, \projname\ is implemented
as a real passive RF monitor, not a simulation.  Second, measured carrier
selection is essential and exposes a methodological artifact in the original
$1.8\,\mathrm{GHz}$ channel.  Third, the main record-level detector achieves
high AUC and low false-positive operation on the main corpus.  Fourth, the same
pipeline remains effective after re-collecting at the measured
$(80,800)\,\mathrm{MHz}$ carrier pair.  Finally, the negative and stress
results are consistent with the design: broad open-set skill recognition is
fragile, while separable skill evidence and fine-window attack evidence provide
the reliable basis for workflow-integrity monitoring.

\section{Related Work}
\label{sec:related}

To contextualize \projname's contributions, we systematically review the literature across host-based auditing, LLM agent security, and physical side-channel analysis. Table~\ref{tab:related_comparison} summarizes this landscape, highlighting the structural gaps that \projname\ addresses.

\begin{table*}[t]
  \centering\small
  \renewcommand{\arraystretch}{1.2}
  \caption{Taxonomy of Workflow Integrity and Side-Channel Defenses vs. \projname. (\textbf{OS-Resilient}: Survives kernel/root compromise; \textbf{Granularity}: Target abstraction level; \textbf{OOB}: Out-of-band isolated channel).}
  \label{tab:related_comparison}
  \begin{tabular}{@{}p{3.5cm}p{5.5cm}p{2.5cm}cccc@{}}
    \toprule
    \textbf{Defense Paradigm} & \textbf{Representative Systems} & \textbf{Granularity} & \textbf{OS-Resilient?} & \textbf{Drift-Aware?} & \textbf{OOB?} \\
    \midrule
    \textbf{Host-based Provenance} & HOLMES~\cite{holmes2019sp}, Unicorn~\cite{han2020unicorn}, ProvDetector~\cite{wang2020provdetector}, MAGIC~\cite{magic2021usenix}, Sleuth~\cite{hossain2017sleuth} & System Call / Process & $\times$ & N/A & $\times$ \\
    \textbf{In-Kernel eBPF / Logs} & Kobra~\cite{kobra2023ndss}, NoDoze~\cite{hassan2019nodze}, Log2vec~\cite{log2vec}, PrioTracker~\cite{liu2021priotracker} & Instruction / Event & $\times$ & N/A & $\times$ \\
    \textbf{LLM Agent Guardrails} & ACE~\cite{ace2026}, SAGA~\cite{saga2026}, StruQ~\cite{struq2025}, AgentDojo~\cite{debenedetti2024agentdojo} & Semantic / Prompt & $\times$ & N/A & $\times$ \\
    \textbf{Bit-Level EM Crypto} & RSA/ECC Extraction~\cite{genkin2014rsa, genkin2016ecdsa}, Screaming~\cite{camurati2018screaming}, BlueScream~\cite{wang2024bluescream} & Cryptographic Bit & \checkmark & $\times$ & \checkmark \\
    \textbf{App-Level EM Profiling} & EMMA~\cite{sehatbakhsh2020emma}, Callan et al.~\cite{callan2014practical}, Yilmaz et al.~\cite{yilmaz2019emi} & Application / OS Loop & \checkmark & $\times$ & \checkmark \\
    \midrule
    \textbf{\projname\ (Ours)} & This paper & \textbf{Agent Skill} & \checkmark & \checkmark & \checkmark \\
    \bottomrule
  \end{tabular}
\end{table*}

\subsection{Host-Based Telemetry and Provenance Auditing}
Traditional host-based intrusion detection systems (HIDS) rely on OS-level telemetry to reconstruct causal execution flows. A massive body of literature has explored system-call-based provenance graphs for Advanced Persistent Threat (APT) detection, including seminal works like HOLMES~\cite{holmes2019sp}, Sleuth~\cite{hossain2017sleuth}, Unicorn~\cite{han2020unicorn}, and ProvDetector~\cite{wang2020provdetector}. Recent advancements have integrated graph representation learning (MAGIC~\cite{magic2021usenix}, ThreaTrace~\cite{wang2022threatrace}), causal inference (CausalIL~\cite{chen2023causalil}), and alert triage optimization (NoDoze~\cite{hassan2019nodze}, PrioTracker~\cite{liu2021priotracker}, Nodemerge~\cite{hassan2020nodemerge}) to filter semantic workflow anomalies from massive enterprise logs~\cite{log2vec, hassan2019nodze, zheng2021poirology}. Furthermore, in-kernel observability frameworks utilizing eBPF (e.g., Kobra~\cite{kobra2023ndss}) have become the standard for low-overhead tracing.

Despite their sophistication, these software-layer defenses share a fundamental vulnerability: the \emph{symmetric threat model}. They inherently assume the uncompromised integrity of the host OS and the hypervisor~\cite{zhu2025controlled, zhong2025bootloader}. As demonstrated by robust bootloader fuzzing campaigns~\cite{zhong2025bootloader} and controlled preemption attacks~\cite{zhu2025controlled}, once an adversary escalates privileges to the kernel or runtime environment, provenance records can be selectively forged, and eBPF sensors blinded. \projname\ completely sidesteps this limitation by relocating the monitoring trust anchor outside the host's physical and logical boundaries.

\subsection{LLM Agent Attacks and Defenses}
The transition from passive LLMs to autonomous, tool-wielding agents has catalyzed a novel attack surface. Early offensive research exposed indirect prompt injections~\cite{greshake2023not, perez2022ignore, liu2024promptinjection}, which have rapidly evolved into sophisticated workflow hijacking vectors. Adversaries can now weaponize poisoned Retrieval-Augmented Generation (RAG) architectures (e.g., PoisonedRAG~\cite{poisonedrag2025}, ObliInjection~\cite{oblinjection2026}, RAGPoison~\cite{ragpoison2024}) and malicious tool documentation (e.g., ToolHijacker~\cite{toolhijacker2026}, AgentSmith~\cite{agentsmith2024}) to seamlessly subvert an agent's planning logic without direct API access. 

To mitigate these threats, the defense community has proposed extensive application-layer safeguards. These range from prompt isolation and privilege separation architectures (ACE~\cite{ace2026}, SAGA~\cite{saga2026}, StruQ~\cite{struq2025}) to comprehensive safety benchmarks (AgentDojo~\cite{debenedetti2024agentdojo}). While these guardrails successfully restrict non-privileged logical flaws, they execute entirely within the host's memory space. If the execution environment itself is compromised via a supply-chain vulnerability (e.g., malicious PyPI packages in the toolchain~\cite{xiao2025jbomaudit}), software-level constraints fail. \projname\ addresses this by treating the agent execution as a black box, verifying its structural physical footprint instead of its self-reported logs.

\subsection{Out-of-Band and EM Side-Channel Monitoring}
Physical side-channels have long been investigated for out-of-band monitoring. Classic electromagnetic (EM) security research heavily targets offensive cryptanalysis, achieving bit-level extraction of RSA/ECC keys via near-field probes~\cite{genkin2014rsa, genkin2016ecdsa, camurati2018screaming, wang2024bluescream}. Moving up the abstraction hierarchy, instruction- and application-level fingerprinting frameworks (EMMA~\cite{sehatbakhsh2020emma}, Callan et al.~\cite{callan2014practical}) successfully distinguish among small sets of desktop applications or detect smartphone camera activations~\cite{yilmaz2019emi}. Other physical modalities, such as power line monitoring (WattsUp~\cite{clark2013wattsup}, HardFails~\cite{dessouky2019hardfails}) and acoustic snooping (RefleXnoop~\cite{reflexnoop2024}), similarly exploit hardware physics for security.

\projname\ pioneers a distinct, mid-tier granularity: \emph{skill-level} workflow monitoring. Unlike bit-level cryptanalysis (which targets tightly unrolled mathematical loops) or app-level classification (which identifies monolithic programs), LLM agent skills are seconds-long, compositional, and highly dynamic macroscopic workloads. Translating these noisy, non-stationary RF streams into semantic intent requires overcoming severe thermal drift---a challenge largely unaddressed in short-window crypto-EM literature. By leveraging a drift-aware coarse--fine windowing pipeline on SDRs at $2\,\mathrm{cm}$, \projname\ provides a resilient hardware trust anchor that definitively bridges the gap between analog emanations and autonomous workflow integrity.

\section{Discussion and Limitations}
\label{sec:discussion}

\textbf{The Physics of Skill-Level Observation and Feature Bottlenecks.} Our evaluation demonstrates a sharp granularity gradient in side-channel analysis. Unlike bit-level cryptanalysis or monolithic app-level fingerprinting, LLM agent skills manifest as seconds-long, compositional workloads whose EM envelopes are dominated by structural resource choices (compute vs. I/O vs. memory). However, our mutual-information diagnosis reveals a fundamental feature-space bottleneck: while architecturally distinct skills exhibit extreme pairwise separability ($F_1 > 0.94$), scaling to a flat 16-class identification hits a plateau (macro-$F_1 \approx 0.146$). This is not a generic detector failure but a physical reality---distinct skills with similar underlying resource constraints generate highly confusable power draws. This physical overlap explicitly necessitates \projname's two-stage design: relying on coarse-fine sequence voting and structural workflow edit-distance comparisons rather than brittle, single-shot multi-class predictions. Future work may lift this ceiling by exploring contrastive pretraining objectives or phase-structure features that inherently subtract cycle-level nuisance variation.

\textbf{Adversarial Adaptation and Asymmetric Defense.} A sophisticated adversary aware of \projname\ might attempt several adaptive evasion strategies. The attacker could shape a malicious payload to mimic the EM profile of a benign skill, execute payloads in sub-window bursts ($< 0.5\,\mathrm{s}$), or exploit dynamic voltage and frequency scaling (DVFS) manipulation to erase the targeted harmonic footprint. Against EM-mimicry, \projname's confusability-weighted edit distance already penalizes unlikely structural sequences, though detecting perfectly crafted identical-EM payloads remains an open challenge. Against DVFS manipulation, our in-situ band selection methodology explicitly abandons highly volatile CPU clock harmonics in favor of stable, low-frequency PMIC and DRAM fundamentals. Furthermore, if an attacker resorts to deploying host-controlled GPIO/PWM emitters to actively jam the in-band RF spectrum, a hardened \projname\ deployment can trivially fallback to a spectrum-anomaly monitor that flags the environment as "signal-degraded" rather than emitting false-clean verdicts, maintaining the asymmetric defense advantage.

\textbf{Deployment Practicality and Generalization Limits.} Operating via passive SDRs at a $2\,\mathrm{cm}$ standoff cleanly matches realistic deployment constraints in shared enterprise environments (e.g., rack-mounted industrial gateways), entirely preserving the out-of-band trust anchor. With a hardware Bill of Materials (BoM) around $\$1{,}500$, batched inference latency of $0.15\,\mathrm{ms}$, and an online feature extraction pipeline that shrinks raw IQ storage to $\sim\!50\,\mathrm{GB}$/day, the system is highly practical for real-time SOC auditing. However, generalizing physical-layer models remains a recognized challenge. While \projname\ successfully mitigates within-session thermal drift via dynamic polynomial detrending, and proves remarkably robust across carrier frequencies (the $88.3\%$ new-bands replication), cross-device (e.g., different ARM SoCs) and cross-session generalization (e.g., across distinct days with altered ambient interference) without re-calibration remain limited. Transitioning from the current technical-feasibility single-device study to a multi-DUT, multi-environment deployed system requires lightweight, automated deployment-time calibration protocols, which we earmark as the primary trajectory for future research.

\section{Conclusion}
\label{sec:conclusion}

As LLM-driven agents evolve into autonomous, task-driven infrastructure, securing their execution against workflow hijacking requires transcending the symmetric trust boundaries of host-internal software. When adversaries can achieve arbitrary code execution and compromise the underlying operating system, traditional software telemetry---such as system calls, audit logs, and eBPF monitors---can be fundamentally blinded or forged. To address this, we introduced \projname, the first out-of-band, physical-layer integrity monitor for agent workflows. By capturing unintentional electromagnetic (EM) emanations via passive SDRs at $2\,\mathrm{cm}$, \projname\ grounds its security guarantees in unforgeable hardware physics rather than vulnerable host software. To bridge the semantic gap between continuous analog signals and discrete agent intent, we engineered a drift-resistant, event-aware coarse--fine windowing architecture that translates macroscopic hardware execution envelopes into deterministic security verdicts.

Our extensive evaluation, underpinned by a $7.82\,\mathrm{TB}$ RF corpus spanning 38 benign and attack skills, demonstrates the profound efficacy of this approach. \projname\ achieves a production-split ROC AUC of $0.9945$ and a $100\%$ true positive rate at just $1.16\%$ false positive rate, delivering structural integrity validations with a median inference latency of a mere $18\,\mathrm{ms}$. Beyond the system itself, we contribute critical methodological insights to the side-channel community by exposing the measurement pitfalls of OS-governor-driven frequency modulation, establishing a robust, OS-agnostic band-selection standard. By formalizing and successfully validating skill-level EM monitoring, \projname\ establishes a practical, host-independent hardware trust anchor, opening a new frontier for securing the next generation of autonomous AI platforms.

\bibliographystyle{IEEEtran}
\bibliography{reference}

\appendix

\begin{newblk}
\section{Boundary Stress Test for Broad Attack-Skill Recognition}
\label{app:stress}

This appendix supports the robustness discussion in \S\ref{sec:eval:robustness}.
It records a deliberately difficult stress campaign and explains why
\projname\ is evaluated as a staged physical-consistency monitor rather than
as a universal single-shot $22$-class skill recognizer.

\paragraph{Corpus and protocol.}
The \texttt{openclaw\_attack\_v1} campaign exercises the full $22$-skill
attack catalog on a separate host and a third carrier pair
$(248,800)\,\mathrm{MHz}$.  This corpus is distinct from both the headline
laptop corpus and the new-bands replication corpus in Table~\ref{tab:datasets}.
It is therefore used only as a boundary stress test: it mixes host shift, run
shift, carrier shift, and small per-class support.

Twenty-two attack-related skills were ported: $11$ skills with prefix
\texttt{attack\_third\_party\_}, $10$ skills with prefix
\texttt{attack\_bad\_tool\_result\_}, and one idle anchor.  Each attack skill
is a one-to-one reimplementation of a corresponding
\texttt{m4p1e/agent-sentinel} attack entry and emits
\texttt{attack\_begin}, \texttt{payload\_start}, \texttt{payload\_end}, and
\texttt{attack\_end} events for offline alignment.  The usable feature set
contains $155$ records across all $22$ classes.

\paragraph{Negative result.}
A flat $22$-class classifier does not recover reliable skill identity at this
sample budget.  Table~\ref{app:tab:stress-summary} summarizes the evidence.
The combined A+B point estimate, macro-$F_1 = 0.0818$, is only modestly above
random, and its bootstrap confidence interval overlaps the random baseline.
Collapsing the label space to $3$ or $6$ classes does not solve the problem:
the apparent numerical gain is explained by the corresponding rise in the
random baseline.  The most important failure mode is distribution shift:
training on A cycles and testing on B's cycle~5 collapses to $0$ macro-$F_1$.

\begin{table}[t]
  \centering\small
  \caption{Key evidence from the \texttt{openclaw\_attack\_v1} stress campaign.}
  \label{app:tab:stress-summary}
  \begin{tabular}{p{0.25\linewidth}p{0.66\linewidth}}
    \toprule
    Test & Finding \\
    \midrule
    Flat $22$-class & macro-$F_1 = 0.0818$; bootstrap CI $[0.043,0.114]$ overlaps random baseline \\
    Class collapse & $3$-class and $6$-class settings remain at-random after baseline correction \\
    Cross-run transfer & A-only training to B testing reaches $0$ macro-$F_1$ in E9-2 \\
    Feature drift & $26/605$ features have $|d|\geq 0.8$; max $|d| = 6.7$ \\
    Channel ablation & Single-channel and cross-channel-only models are at or below random \\
    Per-class behavior & $14/22$ classes have $F_1 = 0$; the apparent \texttt{stat} bright spot vanishes in a binary test \\
    \bottomrule
  \end{tabular}
\end{table}

\paragraph{Implication for the design.}
This stress campaign supports two design choices made in \S\ref{sec:design}.
First, \projname\ should not be claimed as a universal open-set $22$-class
skill recognizer under small per-class support.  Second, drift-aware evaluation
is necessary: LOCO splitting, cycle-local normalization, temperature
detrending, and training-fold-only feature selection are safeguards against
collection-cycle leakage.  The robust operating setting is therefore the one
evaluated in the body: use skill evidence where physical separability is
strong, and use fine-window attack evidence for localized workflow-hijacking
activity.

\end{newblk}

\begin{newblk}
\section{Measured Carrier Selection and New-Bands Collection}
\label{app:bands}

This appendix supports C1 and C2 in \S\ref{sec:eval}: the system is a real
passive RF prototype, and carrier selection must be measured rather than
assumed from nominal hardware frequencies.

\paragraph{Band-survey iterations.}
The first survey iteration used the default \texttt{ondemand} governor and a
single-threaded random-walk memory workload.  Two pathologies appeared.  First,
the $1.8\,\mathrm{GHz}$ CPU-clock region was partly driven by governor activity.
Second, the memory workload did not create a clear LPDDR4 signature.

The second iteration pinned the host with the \texttt{userspace} governor and
replaced the memory workload with a large in-place radix-sort workload.  This
produced visible CPU/RAM separation below $500\,\mathrm{MHz}$ and around
$700$--$900\,\mathrm{MHz}$.  The memory-heavy signature increased from roughly
$+4.3\,\mathrm{dB}$ to $+7.1\,\mathrm{dB}$ at the best window.  The LPDDR4 zoom
further showed that the RAM-heavy trace is elevated by roughly
$3$--$4\,\mathrm{dB}$ across $700$--$900\,\mathrm{MHz}$ and peaks near
$800\,\mathrm{MHz}$.  By contrast, the $1.8\,\mathrm{GHz}$ region remains
tightly bundled under explicit governor pinning.  This supports the
methodological claim that the legacy $1.8\,\mathrm{GHz}$ channel partly reflects
governor-sensitive activity-frequency modulation rather than a stable
idle-vs.-busy hardware-power signature.

\begin{figure}[t]
  \centering
  \includegraphics[width=\linewidth]{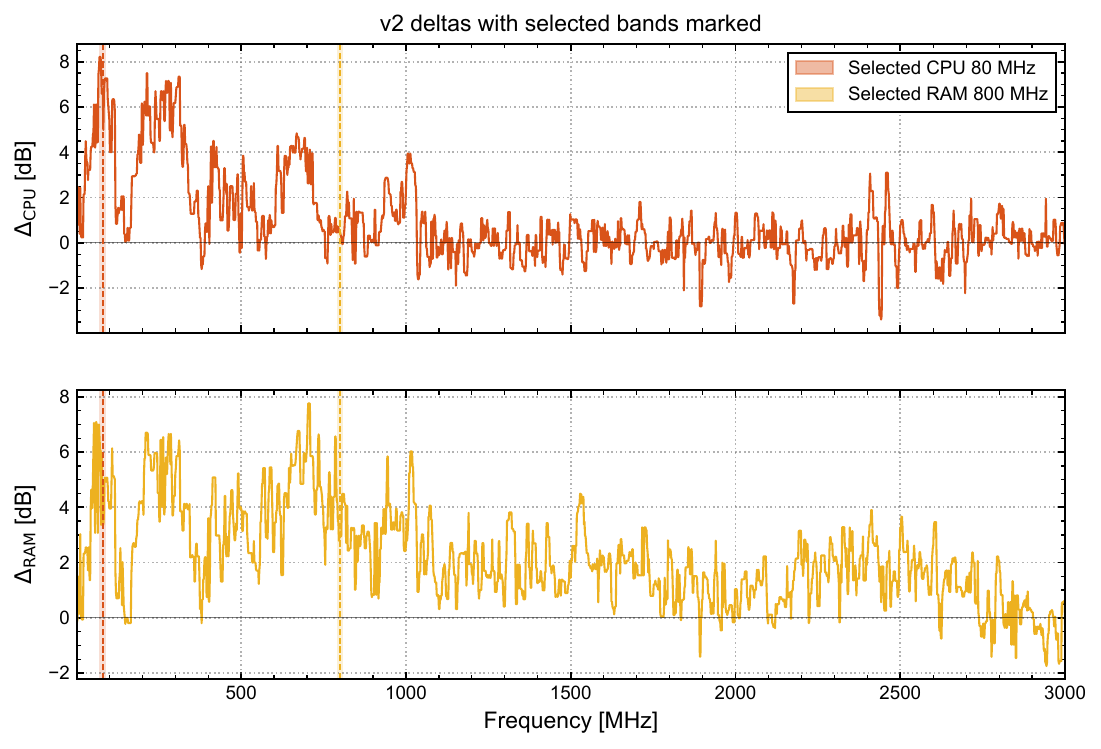}
  \vspace{-0.8em}
  \caption{Band survey v2 workload deltas with the selected CPU/RAM bands highlighted.}
  \label{app:fig:v2-deltas}
  \vspace{-0.8em}
\end{figure}

\begin{figure}[t]
  \centering
  \includegraphics[width=\linewidth]{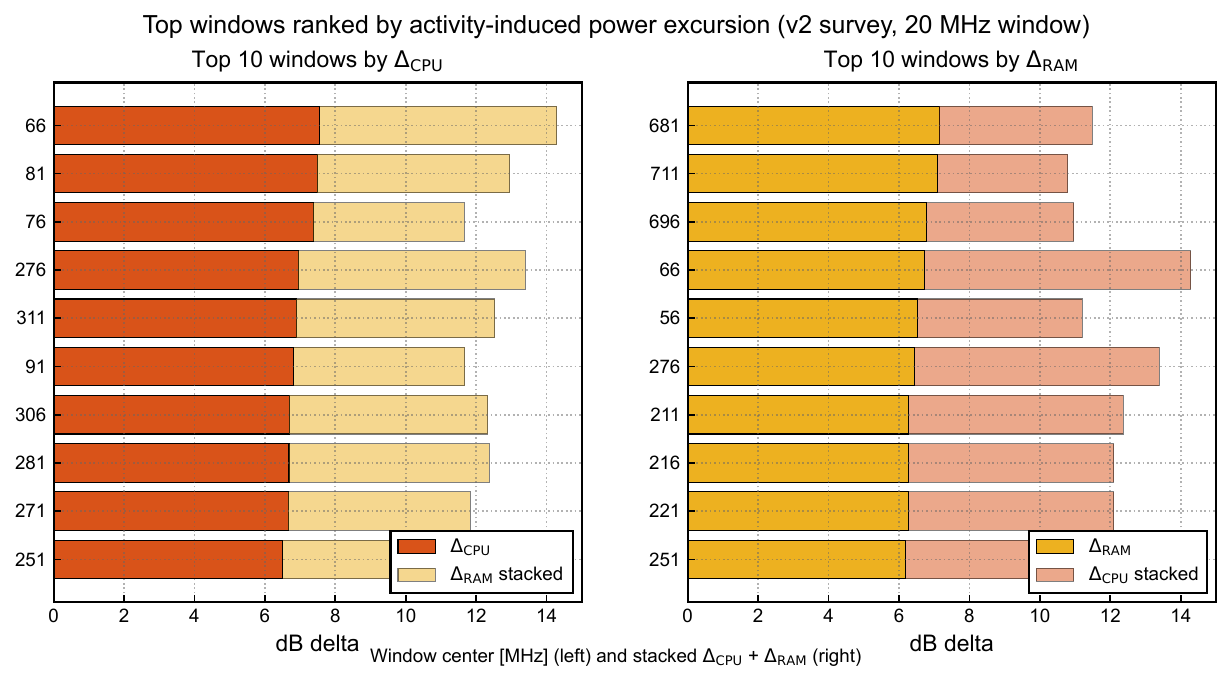}
  \vspace{-0.8em}
  \caption{Top candidate windows ranked by CPU and RAM workload deltas.}
  \label{app:fig:top-windows}
  \vspace{-0.8em}
\end{figure}

\paragraph{New-bands corpus.}
After selecting the $(80,800)\,\mathrm{MHz}$ carrier pair, we recollected a
complete attack benchmark.  A $2\,\mathrm{s}$ pre-flight on each HackRF unit
confirmed clean tuning before the full run.  The full $10$-cycle session ran
for $98$ minutes and produced $550$ IQ files, corresponding to
$440\,\mathrm{GB}$ of raw IQ, together with $221$ per-skill event files and
$11{,}114$ temperature samples.  The session exited cleanly.  Per-cycle
progress was steady, per-skill counts remained balanced across the two
channels, and the two receiver channels exhibited stable amplitude regimes over
the run.

The new-bands replication reaches $83.6\%$ sub-window accuracy,
$88.3\%$ record-vote accuracy, and $90.3\%$ record-level attack recall on the
surviving attack-class subset.  These results support cross-carrier feasibility
while also bounding the claim: residual drift remains, especially in folds
where background windows are conservatively classified as attack.

\end{newblk}

\begin{newblk}
\section{Supplementary Evaluation Details}
\label{app:eval-detail}

This appendix supports C2--C4 in \S\ref{sec:eval}.  It preserves the additional
analyses that substantiate the body claims: skill separability is structured
but not universal; the coarse--fine detector improves localized attack
detection; the new-bands result is useful but still affected by drift; and
runtime overhead is small relative to skill duration.

\paragraph{Skill separability.}
The aggregate $16$-class result on \texttt{big48} is low
(macro-$F_1 = 0.146$), but the pairwise structure is informative.  Among
$120$ pairwise $2$-class tests, $12$ pairs exceed $0.80$ $F_1$, while $37$
pairs fall below $0.55$.  The most separable skill is
\texttt{build\_release\_pipeline}, which reaches $0.956$ against
\texttt{log\_rotate\_compress}, $0.947$ against
\texttt{sensor\_polling\_iot}, and $0.941$ against
\texttt{system\_maintenance}.  The least separable pair is
\texttt{db\_analytics} versus \texttt{video\_streaming} at $0.435$.
The best operational meta-class triples, such as background/build/sensor,
achieve macro-$F_1\in[0.74,0.79]$ under LOCO.  This supports the staged design:
skill-level EM evidence is useful where the physical classes are separable,
but broad open-set recognition remains fragile.

\begin{figure}[t]
  \centering
  \includegraphics[width=\linewidth]{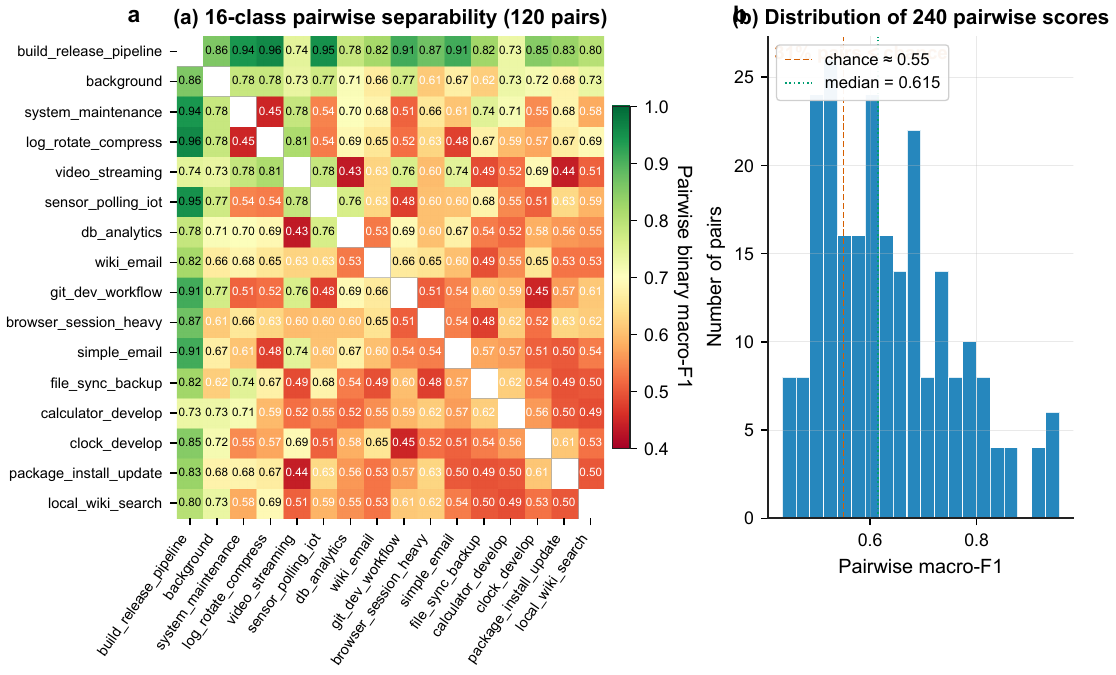}
  \vspace{-0.8em}
  \caption{Pairwise $2$-class separability on \texttt{big48}.}
  \label{app:fig:pairwise}
  \vspace{-0.8em}
\end{figure}

\paragraph{Ablations.}
On \texttt{focused3}, the top-$k$ feature sweep peaks at $k = 65$ with
macro-$F_1 = 0.8977$ and degrades beyond $k = 80$.  On \texttt{big48},
$k = 40$ marginally outperforms $k = 80$, confirming that feature-count
selection is task dependent and must be fit on the training fold only.
Temperature detrending also depends on sample support.  On \texttt{focused3},
a first-order detrend improves over no detrend by $0.018$ macro-$F_1$ and over
a second-order detrend by $0.045$, while smaller datasets can be harmed by
detrending.  Adding raw temperature as a feature changes macro-$F_1$ by less
than $0.005$.  Raw random-forest scores are already well calibrated
(Brier score $0.011$, ECE $0.009$), and post-hoc isotonic recalibration
degrades both metrics.

Band attribution confirms that the headline legacy model uses RF structure but
also motivates the new-bands replication.  Bands $30$--$34$, corresponding to
the legacy $1.8\,\mathrm{GHz}$ region, carry most cumulative importance; masking
this cluster drops AUC substantially.  Cycle leakage is also substantial: many
V10 features have higher mutual information with cycle index than with skill
class, which motivates LOCO evaluation and leakage-controlled preprocessing.

\begin{figure}[t]
  \centering
  \includegraphics[width=\linewidth]{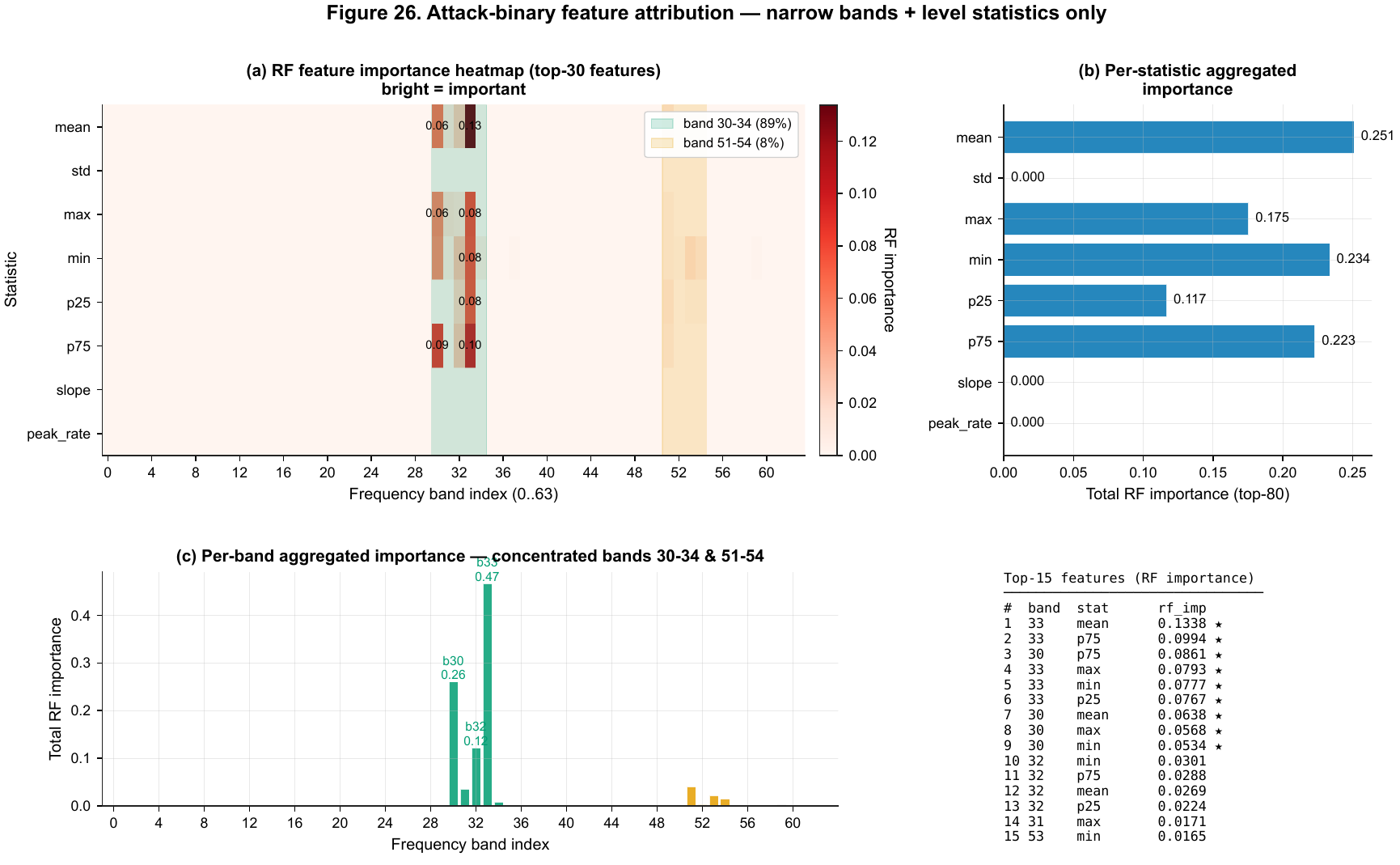}
  \vspace{-0.8em}
  \caption{Random-forest band-level feature attribution on the headline corpus.}
  \label{app:fig:feature-attrib}
  \vspace{-0.8em}
\end{figure}

\paragraph{Coarse--fine attack detection.}
Pilot~A captures $30$ tasks in which the planner is induced to call
\texttt{attack\_third\_party\_rm}.  A flat $2$-class within-session classifier
on $20\,\mathrm{s}$ traces reaches macro-$F_1 = 0.670$ with attack recall
$0.55$.  The coarse--fine stage-2 detector improves the record-level result to
$0.9398$ accuracy with attack recall $0.833$.

The larger attack pilot exercises the full $22$-skill attack catalog over
$10$ cycles and $299$ records.  Across four stage-2 configurations,
record-level accuracy remains at least $0.9252$ and attack recall lies between
$0.833$ and $0.861$, as shown in Table~\ref{app:tab:cf-detail}.  This supports
the body claim in \S\ref{sec:eval:detection}: fine-window evidence exposes
short malicious payloads that are diluted in whole-record representations.

\begin{table}[t]
  \centering\small
  \caption{Coarse--fine $3$-class attack-detection results on
  \texttt{big48\_chunk1}+\texttt{20260423b}.}
  \label{app:tab:cf-detail}
  \begin{tabular}{lcccc}
    \toprule
    Config & Sub-win acc. & Record acc. & bg recall & atk recall \\
    \midrule
    v1 & $0.816$ & $\mathbf{0.9398}$ & $0.250$ & $0.833$ \\
    v2 & $0.787$ & $0.9252$ & $0.139$ & $0.844$ \\
    v3 & $0.809$ & $0.9398$ & $0.222$ & $0.861$ \\
    no-temp & $0.809$ & $0.9398$ & $0.222$ & $0.861$ \\
    \bottomrule
  \end{tabular}
\end{table}

\paragraph{New-bands stability.}
Figures~\ref{app:fig:newbands-confusions} and
\ref{app:fig:newbands-percycle} expand the new-bands replication:
cycles $3$, $7$, and $8$ reach $100\%$ record-vote accuracy, while cycles
$1$, $5$, $6$, and $10$ are lower.  In cycles $1$, $6$, and $10$, every
background sub-window is classified as attack, explaining much of the per-fold
variance.  This does not invalidate the detection claim, but it bounds it: the
new-bands corpus supports cross-carrier feasibility, while a larger or
temperature-stratified recollection is needed to reduce false alerts under
new-band drift.

\begin{figure}[t]
  \centering
  \begin{subfigure}[t]{0.48\linewidth}
    \centering
    \includegraphics[width=\linewidth]{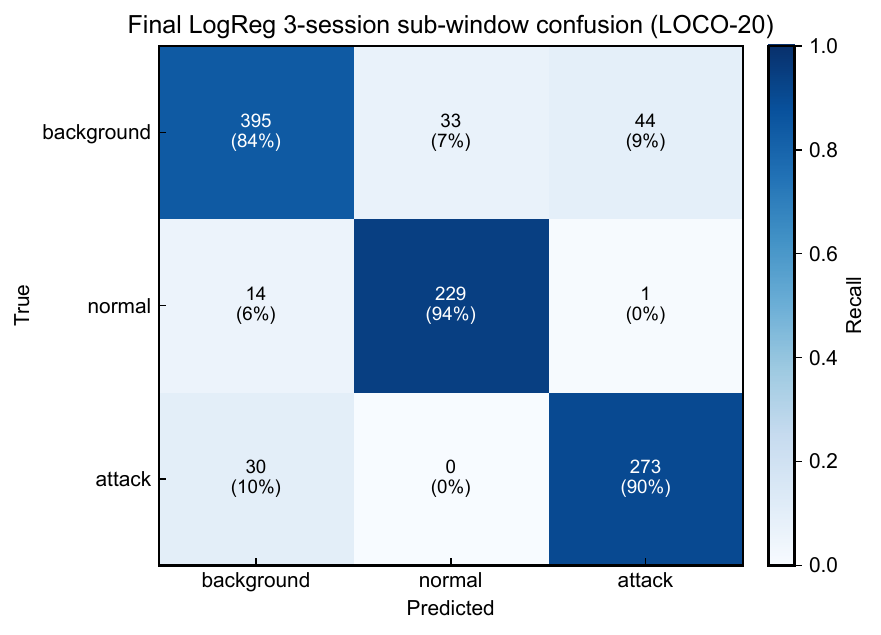}
    \caption{Sub-window pooled.}
    \label{app:fig:newbands-conf-subwindow}
  \end{subfigure}
  \hfill
  \begin{subfigure}[t]{0.48\linewidth}
    \centering
    \includegraphics[width=\linewidth]{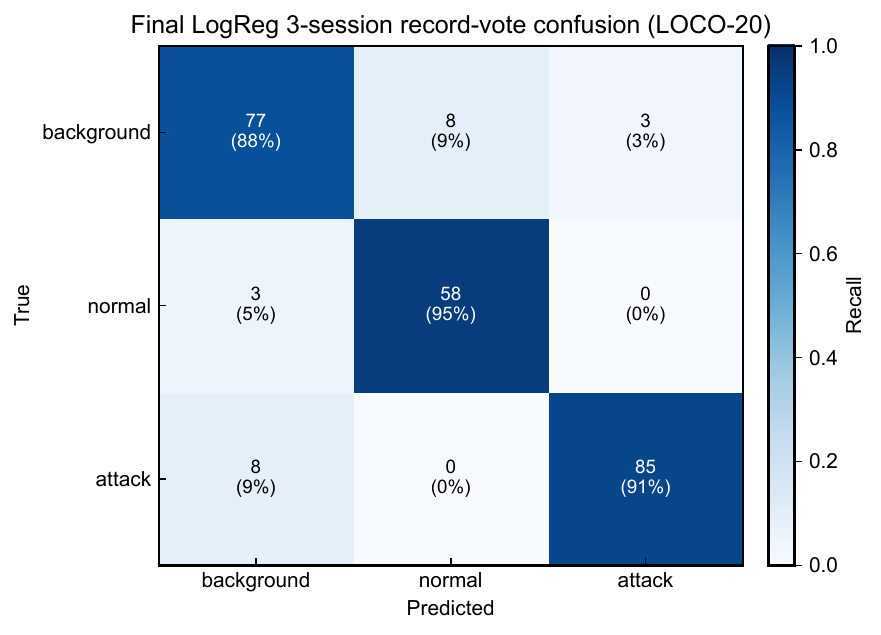}
    \caption{Record vote.}
    \label{app:fig:newbands-conf-recordvote}
  \end{subfigure}
  \vspace{-0.8em}
  \caption{Confusion matrices for the new-bands LOCO experiment.}
  \label{app:fig:newbands-confusions}
  \vspace{-0.8em}
\end{figure}

\begin{figure}[t]
  \centering
  \includegraphics[width=\linewidth]{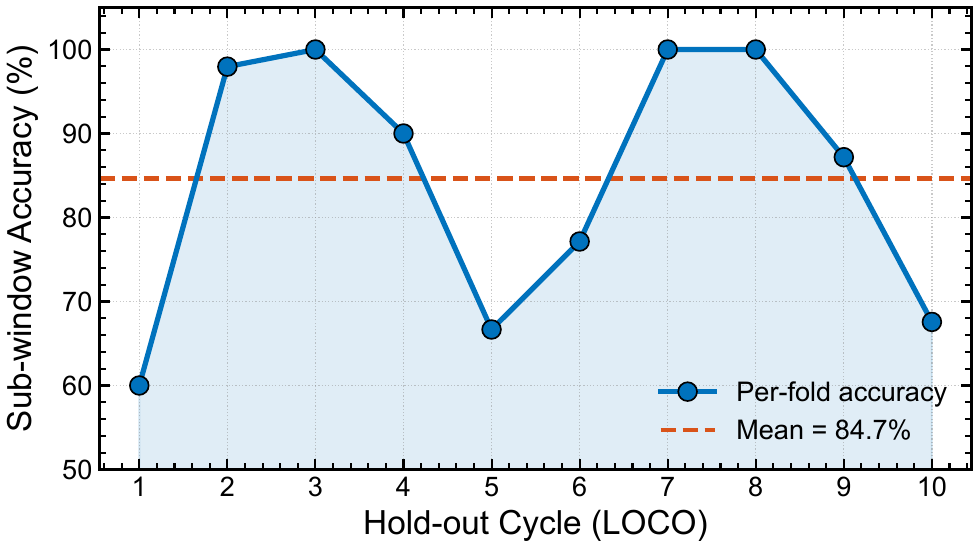}
  \vspace{-0.8em}
  \caption{Per-fold accuracy across the $10$ new-bands LOCO folds.}
  \label{app:fig:newbands-percycle}
  \vspace{-0.8em}
\end{figure}

\paragraph{Baselines, failure modes, and runtime.}
A naive cross-corpus binary attack classifier is not sufficient.  Training a
flat $2$-class model with \texttt{big48} normals as benign and the $22$-class
attack corpus as malicious yields apparent accuracy $0.993$, but
macro-$F_1 = 0.498$ and catches zero of $11$ attack records.  One-class anomaly
baselines also remain near chance, with AUC in $[0.44,0.58]$.  These failures
support the design choice to use supervised fine-window attack evidence rather
than density anomaly alone.

Post-feature inference is not the bottleneck.  The median per-record inference
latency is $18\,\mathrm{ms}$, the $p_{99}$ latency is $29\,\mathrm{ms}$, and
batched prediction amortizes to approximately $0.15\,\mathrm{ms}$ per record.
Because agent skills last seconds and fine windows last hundreds of
milliseconds, sensing and windowing dominate wall-clock delay.

\begin{figure}[t]
  \centering
  \includegraphics[width=\linewidth]{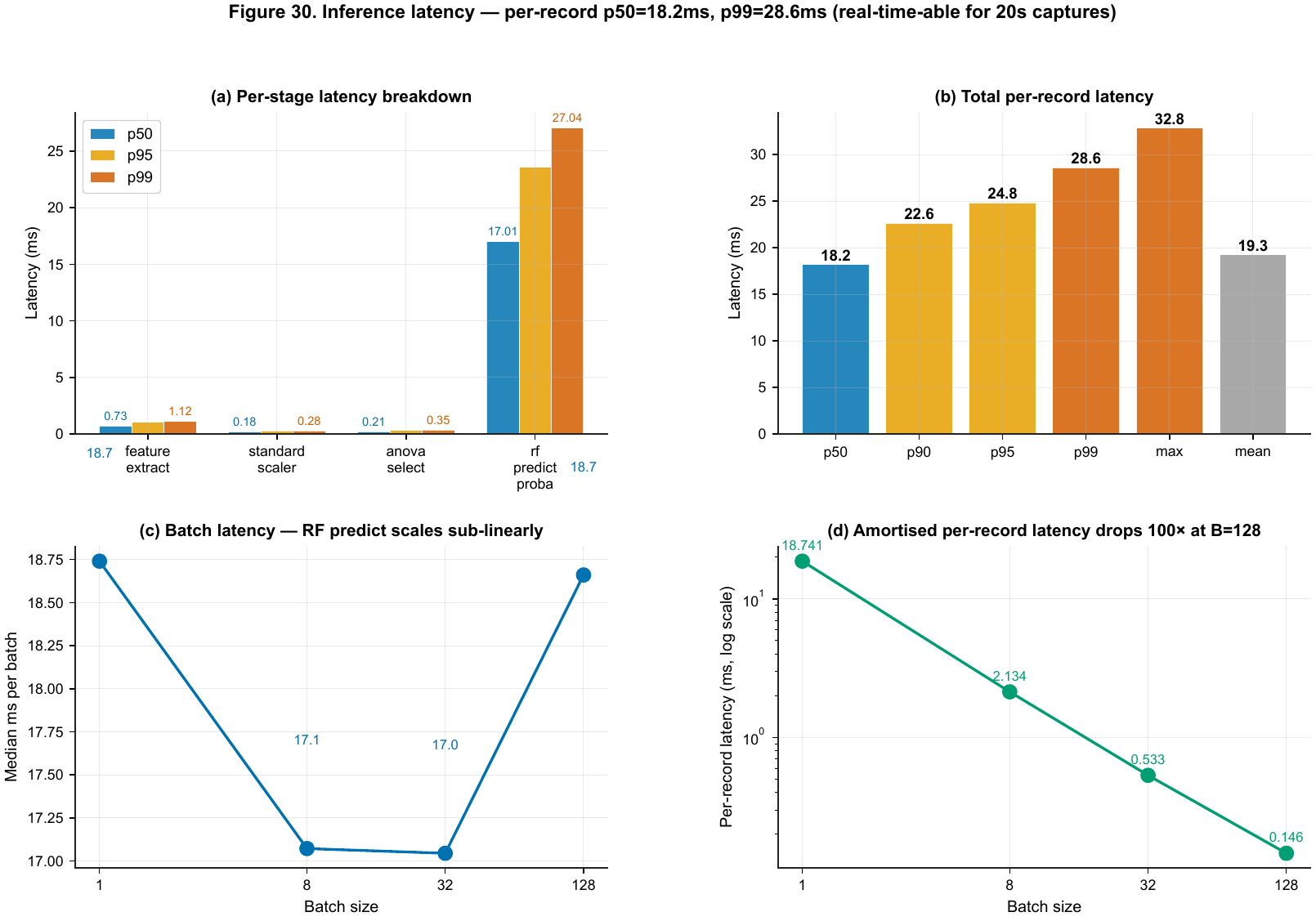}
  \vspace{-0.8em}
  \caption{Per-record inference latency breakdown.}
  \label{app:fig:latency-detail}
  \vspace{-0.8em}
\end{figure}

\end{newblk}

\end{document}